\documentclass[10pt, final, journal, twocolumn, letterpaper]{IEEEtran} 

\usepackage[latin1]{inputenc}
\usepackage{xcolor}
\usepackage{cite}
\usepackage{amssymb}
\usepackage{amsthm,amsmath,array}
\usepackage{graphicx}                  
\usepackage{rotating}
\usepackage{times}
\usepackage{bm}
\usepackage{enumitem}
\usepackage{epsfig,psfrag}
\usepackage{subcaption}
\usepackage[font=small]{caption}
\usepackage[export]{adjustbox}
\usepackage{winsnotation}
\usepackage{tabu}
\usepackage{url}

\usepackage{tabularx}
\newcolumntype{L}[1]{>{\raggedright\arraybackslash}p{#1}}
\newcolumntype{C}[1]{>{\centering\arraybackslash}p{#1}}
\newcolumntype{R}[1]{>{\raggedleft\arraybackslash}p{#1}}

\definecolor{darkgreen}{rgb}{0.0, 0.5, 0.0}
\definecolor{magenta_}{rgb}{1.0, 0, 1.0}

\newcommand{\be}{\begin{equation}}
\newcommand{\ee}{\end{equation}}
\newcommand{\ist}{\hspace*{.3mm}}
\newcommand{\rmv}{\hspace*{-.3mm}}

\newcommand{\nn}{\nonumber}

\definecolor{red_plot}{rgb}{0.66,0,0}
\definecolor{blue_plot}{rgb}{0,0.3,0.7}
\makeatletter
\newcommand*\bigcdot{\mathpalette\bigcdot@{.5}}
\newcommand*\bigcdot@[2]{\mathbin{\vcenter{\hbox{\scalebox{#2}{$\m@th#1\bullet$}}}}}
\makeatother
\usepackage{mathtools}

\newcommand{\anchoridxsym}{j}               
\newcommand{\anchoridx}{^{(\anchoridxsym)}} 

\newcommand{\mpcidxsym}{k}          

\newcommand{\timestepsym}{n}            

\newcommand{\vu}[2]{\mbox{$#1\,\text{#2}$}} 

\newcommand{\diag}[1]{\mathrm{diag}}   

\newcommand{\exportFigures}{false}
\ifthenelse{\equal{\exportFigures}{true}}
{
  \usepgfplotslibrary{external}\tikzexternalize[prefix=tikz/]
}
{}

\newcommand{\pgfref}[1]
{\ifthenelse{\equal{\exportFigures}{true}}
{\tikzexternaldisable\ref{#1}\tikzexternalenable}
{\ref{#1}}}

\newcommand{\trans}{^\text{T}}

\IEEEoverridecommandlockouts

\begin{document}
\allowdisplaybreaks
\frenchspacing

\title{A Belief Propagation Algorithm\\ for Multipath-Based SLAM}

\author{\normalsize Erik Leitinger,~\IEEEmembership{\normalsize Member,~IEEE}, Florian Meyer,~\IEEEmembership{\normalsize Member,~IEEE}, 
Franz Hlawatsch,~\IEEEmembership{\normalsize Fellow,~IEEE},\\
Klaus Witrisal,~\IEEEmembership{\normalsize Member, IEEE}, 
Fredrik Tufvesson, \IEEEmembership{\normalsize Fellow, IEEE}, and
Moe Z. Win, \IEEEmembership{\normalsize Fellow, IEEE}
\thanks{ E.\ Leitinger and K.\ Witrisal are with the Signal Processing and Speech Communication Laboratory, Graz University of Technology, Graz, Austria, (e-mail: \{erik.leitinger, witrisal\}@tugraz.at).
F.\ Meyer and M.\ Z.\ Win are with the Laboratory for Information and Decision Systems, Massachusetts Institute of Technology, Cambridge, MA, USA (e-mail: \{fmeyer, moewin\}@mit.edu). 
F.\ Hlawatsch is with the Institute of Telecommunications, TU Wien, Vienna, Austria and with Brno University of Technology, Brno, Czech Republic   
(e-mail: franz.hlawatsch@tuwien.ac.at).
F.\ Tufvesson is with the Department of Electrical and Information Technology, Lund University, Lund, Sweden (e-mail: fredrik.tufvesson@eit.lth.se). 
This work was supported by the Austrian Science Fund (FWF) under grants J 4027, J 3886, P 27370-N30, and P 32055-N31; 
by the U.S.\ Department of Commerce, National  Institute of Standards and Technology under Grant 70NANB17H17; and by the Czech Science Foundation (GA\v{C}R) 
under grant 17-19638S. Parts of this paper were presented at IEEE ICC, ANLN Workshop 2017, Paris, France, June 2017.
}}

\maketitle

\begin{abstract}
  We present a simultaneous localization and mapping (SLAM) algorithm that is based on radio signals and the association of specular multipath components (MPCs) with geometric features. Especially in indoor scenarios, \textit{robust} localization from radio signals is challenging due to diffuse multipath propagation, unknown MPC-feature association, and limited visibility of features. In our approach, specular reflections at flat surfaces are described in terms of virtual anchors (VAs) that are mirror images of the physical anchors (PAs). The positions of these VAs and possibly also of the PAs are unknown. We develop a Bayesian model of the SLAM problem and represent it by a factor graph, which enables the use of belief propagation (BP) for efficient marginalization of the joint posterior distribution. The resulting BP-based SLAM algorithm 
detects the VAs associated with the PAs and estimates jointly the time-varying position of the mobile agent and the positions of the VAs and possibly also of the PAs, thereby leveraging the MPCs in the radio signal for improved accuracy and robustness of agent localization. The algorithm has a low computational complexity and scales well in all relevant system parameters. Experimental results using both 
synthetic measurements and real ultra-wideband radio signals demonstrate the excellent performance of the algorithm in challenging indoor environments.

\end{abstract}
     

\section{Introduction}\label{sec:Introduction}

The goal of simultaneous localization and mapping (SLAM) \cite{Thrun2005,DurrantWhyte2006} is to estimate the time-varying pose of a mobile agent---including the agent's position---and a map of the surrounding environment, from measurements provided by one or multiple sensors. SLAM is important in many fields including robotics \cite{Thrun2005}, autonomous driving \cite{BressonTIV2017}, location-aware communication \cite{DiTaranto2014SPM}, and robust indoor localization \cite{WitrisalSPM2016, BarConGioWin:L14,DardariTVT2015, GuidiTMC2016,LeitingerGNSS2016}. Achieving a required level of accuracy robustly is still elusive in indoor environments characterized by harsh multipath channel conditions. Therefore, most existing systems supporting multipath channels either use sensing technologies that mitigate multipath effects \cite{WymeerschTC2012} or fuse multiple information sources \cite{ShenJSAC2012,WinSheDai:J18}. 

In multipath-assisted indoor localization \cite{WitrisalSPM2016, GentnerTWC2016,LeitingerGNSS2016, LeitingerJSAC2015, GuidiTMC2016,GuerraSSP2018, ZhangGLOBECOM}, 
the relation of multipath components (MPCs) with the local geometry potentially turns multipath propagation from an impairment into an advantage. This paper presents a SLAM algorithm for robust indoor localization based on radio signals. The radio signals are transmitted from a mobile agent to base stations, called physical anchors (PAs). MPCs due to specular reflections are modeled by virtual anchors (VAs), which are mirror images of the PAs \cite{Borish1984}. Our algorithm detects the VAs associated with the PAs and estimates the VA and possibly also PA positions jointly with the time-varying position of the mobile agent. The algorithm is designed to cope with harsh multipath channel conditions, which tend to lead to measurements with a high level of false alarms and missed detections.  While MPCs can be generated by various propagation phenomena such as specular reflections, scattering, and diffraction, our model focuses on PA/VA-related MPCs; all the other MPCs are modeled as interference, even if they contain geometric information. We note that MPCs associated with scatter points are considered in the feature model used in \cite{GentnerTWC2016}.

\subsection{Feature-based SLAM}\label{sec:introFBSLAM}

The proposed algorithm follows the feature-based approach to SLAM \cite{DurrantWhyte2006, Dissanayake2001, MullaneTR2011}. The map is represented by an unknown number of \textit{features} with unknown spatial positions, whose states are estimated in a sequential (time-recursive) way. In our model, the features are given by the PAs and VAs. Prominent feature-based SLAM algorithms are extended Kalman filter SLAM (EKF-SLAM) \cite{Dissanayake2001}, Rao-Blackwellized (RB)-SLAM (dubbed FastSLAM) \cite{DurrantWhyte2006,MontemerloAAAI2002,GentnerTWC2016}, variational-inference-based SLAM \cite{LundgrenTSP2016, FatemiCAMSAP2015}, and set-based SLAM \cite{MullaneTR2011, DeuschTSP2015, FatemiTSP2017}. Recently, feature-based SLAM methods that exploit position-related information in radio signals were introduced \cite{LeitingerICC2015, GentnerTWC2016, ZhuTufvesson2015, GentnerHindawi2017,GuerraSSP2018, ZhangGLOBECOM}. Most of these methods operate on estimated parameters related to MPCs, such as distances (which are proportional to delays), angles-of-arrival (AoAs), or angles-of-departure (AoDs) \cite{Fleury1999, RichterPhD2005, Shutin2011, SalmiTSP2009, BadiuTSP2017}. These parameters are estimated from the signal in a preprocessing stage and are considered as ``measurements'' by the SLAM method. An important aspect is the \emph{data association} (DA) between these measurements and the PAs or VAs. 

Feature-based SLAM is closely related to multitarget tracking (MTT), and MTT methods have been adapted to feature-based SLAM \cite{MullaneTR2011, DeuschTSP2015, LeitingerICC2017}.
MTT methods that are applicable to SLAM include the joint probabilistic DA (JPDA) filter \cite{BarShalom11} and the joint integrated probabilistic DA (JIPDA) filter \cite{Musicki2004}. An approach similar to the JIPDA filter is taken by the methods presented in \cite{HorridgeFusion2009,Horridge2011PHD}. More recently, the use of belief propagation (BP) \cite{Loeliger2004SPM,KschischangTIT2001} was introduced for probabilistic DA within MTT in \cite{WilliamsTAE2014} and for multisensor MTT in \cite{MeyKroWilLauHlaBraWin:J18,MeyerTSP2017,SoldiTSP2019}. In particular, the BP algorithms in 
\cite{MeyerTSP2017,MeyKroWilLauHlaBraWin:J18,SoldiTSP2019} are based on a factor graph representation of the multisensor MTT problem and have a computational complexity that scales only quadratically in the number of objects (targets) and linearly in the number of sensors. MTT methods that are based on random finite sets and embed a BP algorithm for probabilistic DA were presented in \cite{WilliamsTAES2015, KropfreiterFUSION2016,MeyKroWilLauHlaBraWin:J18}.
We finally note that our approach to feature-based SLAM is also related to multisensor target tracking with uncertain sensor locations using Bayesian methods \cite{MarrsProcAero2001}.

\subsection{Contributions and Organization of the Paper}
\label{sec:contrib_organ}

Here, we propose a BP-based, Bayesian detection and estimation algorithm for SLAM using radio signals. Our algorithm jointly performs probabilistic DA and sequential estimation of the states of a mobile agent and of ``potential features'' (PFs) characterizing the map. We use a probabilistic model for feature existence where each PF state is augmented by a binary existence variable and associated with a \emph{probability of existence}, which is also estimated. The proposed algorithm is inspired by the BP algorithms for multisensor MTT presented in \cite{MeyerTSP2017}, and will hence be briefly referred to as \emph{BP-SLAM} algorithm. Probabilistic DA and state estimation are performed by running BP on a factor graph \cite{Loeliger2004SPM,KschischangTIT2001} representing the statistical structure of the SLAM problem. The BP approach leverages conditional statistical independencies to achieve low complexity and high scalability. In fact, in contrast to conventional SLAM algorithms such as EKF-SLAM and RB-SLAM, the proposed BP-SLAM algorithm assumes that in each time step the agent state and all the feature states are a priori independent.

Our probabilistic model for DA and feature existence uncertainty allows the BP-SLAM algorithm to succeed in the particularly challenging range-only SLAM problem \cite{BlancoICRA2008}, which cannot be addressed straightforwardly by established SLAM techniques \cite{Thrun2005,DurrantWhyte2006}. Performing SLAM only from range measurements is challenging because (i) when new features are initialized, the probability distributions of the PAs/VAs are annularly shaped and thus cannot be well represented by a Gaussian distribution; and (ii) DA is more difficult compared to classical SLAM problems in robotics since PAs/VAs at widely different positions can generate almost identical range measurements. In particular, the algorithm of \cite{BlancoICRA2008} is able to cope with a range-only measurement model, but not with DA uncertainty. To the best of our knowledge, the proposed algorithm---along with a preliminary version presented in \cite{LeitingerICC2017}---is the first BP algorithm for feature-based SLAM with probabilistic DA that is also suitable for range-based SLAM. 

Key innovative contributions of this paper include the following:

\vspace{1.5mm}

\begin{itemize}
  \item We establish a Bayesian model for feature-based SLAM that uses MPC parameters extracted from radio signals as input measurements 
  and models probabilistically the appearance and disappearance of PAs/VAs as well as the DA uncertainty.

\vspace{1.5mm}
  
  \item Based on a factor graph representation of this model, we develop a scalable BP algorithm that estimates the state of the mobile agent and the numbers and positions of PAs/VAs.

\vspace{1.5mm}
  
 \item We evaluate the performance of the proposed algorithm on synthetic and real data. Our experimental results demonstrate the algorithm's high accuracy and robustness.

\vspace{1.5mm}

\end{itemize}

We apply the proposed BP-SLAM algorithm to the challenging setup of a range-only measurement model. However, bearing information (AoA and AoD of MPCs) or information derived from inertial measurement unit sensors can be easily incorporated in the BP-SLAM algorithm, and this would lead to a significant performance gain. For simplicity, we assume time synchronization between the PAs and the mobile agent. However, the BP-SLAM algorithm can be extended to nonsynchronized PA-agent links along the lines of \cite{GentnerTWC2016} (based on the fact that the relevant geometric information is also contained in the time differences of the MPCs \cite{LeitingerJSAC2015,GentnerIPIN2013}) or to joint SLAM and synchronization along the lines of  \cite{etzlinger17}. Furthermore, we assume that the probabilities with which the preliminary signal analysis stage (producing measurements) detects features in the radio signals are known; however, an adaptive extension to unknown and time-varying detection probabilities can be obtained along the lines of \cite{LeitingerICC2017,SoldiTSP2019,LeitingerSSP2018}.

This paper advances over our conference paper \cite{LeitingerICC2017} in that it replaces the heuristic used therein for determining the initial distribution of new PFs by an improved Bayesian technique. Furthermore, the factor graph and BP algorithm of \cite{LeitingerICC2017} are extended by the introduction of ``new PFs,'' i.e., features that generate measurements for the first time. 

The remainder of this paper is organized as follows. 
Section \ref{sec:signalmodel} considers the received radio signals and the MPC parameters. Section \ref{sec:systemmodel} describes the system model and provides a statistical formulation of the SLAM problem. 
The joint posterior distribution of the states and the corresponding factor graph are derived in Section \ref{sec:FGderivation}. In Section \ref{sec:BPmessgepassing}, the proposed BP-SLAM algorithm is presented. The results of numerical experiments are reported in Section \ref{sec:results}. Section \ref{sec:concl} concludes the paper.

\section{Radio Signal and MPC Parameters}\label{sec:signalmodel} 
\begin{figure}[t!]
\vspace*{1mm}
\centering
\includegraphics[scale=1]{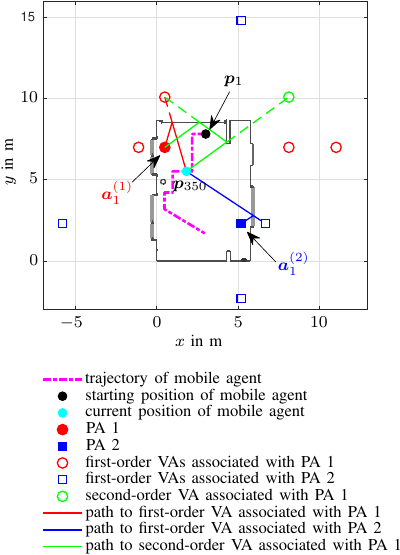}
\vspace{2mm}
\caption{Example of an environment map (floor plan). The PAs at fixed positions $\V{a}_{1}^{(1)}$ (PA 1) and $\V{a}_{1}^{(2)}$ (PA 2) are indicated by, respectively, a red bullet and a blue box within the floor plan. The magenta dashed-dotted line represents the trajectory of the mobile agent. The starting position of the mobile agent, $\V{p}_1$, and the current position $\V{p}_{350}$ (at discrete time $n \!=\! 350$) are indicated by a black and a cyan bullet, respectively. The red circles and blue squares outside the floor plan indicate some of the geometrically expected first-order VAs associated with PA 1 and PA 2, respectively. The green circle outside the floor plan indicates one second-order VA associated with PA 1. Two exemplary first-order reflection paths between the mobile agent at position $\V{p}_{350}$ and two first-order VAs associated with the two PAs are shown by red and blue lines. One exemplary second-order reflection path between the mobile agent at position $\V{p}_{350}$ and a second-
order VA 
associated with PA 1 is shown by a green line.}
\label{fig:VAgeometry_SLAM} 
\end{figure}

Radio signal based SLAM \cite{GentnerTWC2016, LeitingerICC2015, WitrisalSPM2016, ZhuTufvesson2015} associates specular MPC parameters estimated from the received signals with ``geometrically expected'' parameters,
such as distances (related to delays), AoAs, and AoDs. These parameters are modeled in terms of the positions of the mobile agent and of the PAs or VAs. The VA positions are mirror images of the PA positions that are induced by reflections at flat surfaces---typically walls---and thus depend on the surrounding environment (floor plan) \cite{Borish1984}. For each reflection path, the length from the PA (via the flat surface) to the mobile agent is equal to the length from the VA to the mobile agent. Even though the mobile agent moves, the VAs remain static as long as the PAs and the flat surfaces (floor plan) are static. Note that the VA positions are unknown because the floor plan is unknown. As an example, Fig.~\ref{fig:VAgeometry_SLAM} depicts a floor plan with two PAs and some of the corresponding first-order VAs as well as one second-order VA belonging to some larger flat surfaces. Also shown are three reflection paths (for the two PAs) for mobile agent position $\V{p}_{350}$. The construction of higher-
order VAs is described in \cite[Ch.~2]{MeissnerPhD2014}.

We consider a mobile agent with unknown time-varying position $\V{p}_n \rmv\!\in\rmv \mathbb{R}^2$ and $J$ PAs with possibly unknown positions $\V{a}_{1}^{(j)} \rmv\!\in\rmv \mathbb{R}^2\rmv$, $j \rmv=\rmv 1,\ldots,J$, where $J$ is assumed to be known. Associated with the $j$th PA, there are $L_n^{(j)} \!-\rmv 1$ VAs at unknown positions $\V{a}_{l}^{(j)} \!\rmv\in\rmv \mathbb{R}^2\rmv$, $l \rmv=\rmv 2,\ldots,L_n^{(j)}\!$. The PAs and VAs will also be referred to as \emph{features}. The number of features, $L_n^{(j)}\!$, is unknown and time-varying, and it depends on the agent position $\V{p}_n$. In each discrete time slot $n$, the mobile agent transmits a radio signal $s(t)$ and the PAs act as receivers. However, the proposed algorithm can be easily reformulated for the case where the PAs act as transmitters and the mobile agent acts as a receiver. The baseband signal received by the $j$th PA is modeled as \cite{LeitingerJSAC2015}
\begin{align}
\nn\\[-6mm]
\label{eq:rx_signal}
  r_n^{(j)}(t) \ist= \sum_{l=1}^{L_n^{(j)}}w_{l,n}^{(j)} \ist s\big(t \rmv-\rmv \tau_{l,n}^{(j)}\big) +\ist d_{n}^{(j)}(t) \ist+\ist n_{\text{AWGN}}(t)\ist.
\\[-6mm]\nn
\end{align}
Here, the first term on the right-hand side describes the contribution of $L_n^{(j)}$ specular MPCs with complex amplitudes $w_{l,n}^{(j)}$ and delays $\tau_{l,n}^{(j)}$, where $l \rmv\in\rmv \Set{L}_n^{(j)} \!\rmv\triangleq\rmv \big\{ 1,\dots,L_n^{(j)}\big\}$. These MPCs correspond to features (PAs or VAs). The delays $\tau_{l,n}^{(j)}$ are proportional to the distances (ranges) between the agent and 
either the $j$th PA (for $l \!=\! 1$) or the associated VAs (for $l \rmv\in\rmv \{ 2,\dots,L_n^{(j)}\}$). That is, $\tau_{l,n}^{(j)} \rmv=\rmv \big\|\V{p}_n \!-\rmv \V{a}_{l}^{(j)}\big\|\big/c$, where $c$ is the speed of light. The second term in \eqref{eq:rx_signal}, $d_{n}^{(j)}(t)$, represents the diffuse multipath component, which interferes with the specular MPC term. The third term in \eqref{eq:rx_signal}, $n_{\text{AWGN}}(t)$, is additive white Gaussian noise. We note that expression \eqref{eq:rx_signal} presupposes a common time reference at the mobile agent and at the PAs. However, our algorithm can be extended to an unsynchronized system along the lines of \cite{GentnerTWC2016}, exploiting the fact that the relevant geometric information is preserved in the time differences of the MPC delays \cite{LeitingerJSAC2015}, \cite{GentnerIPIN2013}. Furthermore, our algorithm can also be extended to the case where the MPC parameters include AoAs and/or AoDs in addition to the delays $\tau_{l,n}^{(j)}$. An 
extension that uses the complex MPC amplitudes $w_{l,n}^{(j)}$ to directly estimate the detection probability of each feature has been proposed in \cite{LeitingerSSP2018}.

In each time slot $n$ and for each PA $j \rmv\in\rmv \{1,\ldots,J\}$, a parametric radio channel estimator \cite{Fleury1999, RichterPhD2005, Shutin2011, BadiuTSP2017,SalmiTSP2009,JostTAP2012} processes the radio signals $r_n^{(j)}(t)$ and produces $M_n^{(j)}$ MPC parameter estimates along with estimates of the corresponding complex amplitudes, with $m \rmv\in\rmv \Set{M}_n^{(j)} \!\triangleq\rmv \{ 1,\dots,M_n^{(j)} \}$. The set of estimated MPC parameters $\Set{M}_n^{(j)}$ is related to the set of specular MPCs $\Set{L}_n^{(j)}$ as follows. It is possible that some specular MPCs are not ``detected'' by the radio channel estimator and thus do not produce an MPC parameter estimate, and it is also possible that some estimates do not correspond to MPCs. Accordingly, $M^{(j)}_n = \big|\Set{M}_n^{(j)}\big|$ may be smaller than, equal to, or larger than $L^{(j)}_n = \big|\Set{L}_n^{(j)}\big|$. (Here, $|\cdot|$ denotes the cardinality of a set.) Note also that $M_n^{(j)}$ depends on the agent position $\V{p}_n$ and 
on the environment. The amplitude estimates are used to calculate estimates of the MPC parameter variances, if such estimates are not provided by the channel estimator (see Section~\ref{sec:sim_param:real}). We denote by $\V{z}_{m,n}^{(j)}$ with $m \rmv\in\rmv \Set{M}_n^{(j)}$ the estimated parameters of the $m$th MPC of PA $j$. The stacked vectors $\V{z}_n^{(j)} \!\triangleq \big[\V{z}_{1,n}^{(j)\text{T}} \cdots\ist \V{z}_{M_n^{(j)}\!,n}^{(j)\text{T}} \big]\trans\rmv$ are used as noisy ``measurements'' by the proposed BP-SLAM algorithm.

\section{System Model and Statistical Formulation}\label{sec:systemmodel}
\subsection{Agent State and PF States}
\label{sec:vec_description}

The state of the mobile agent at time $n$ is $\V{x}_n \!\triangleq\rmv [\V{p}_n\trans \; \V{v}_n\trans]\trans\rmv$, where $\V{v}_n$ is the agent's velocity. The PFs are indexed by the tuple $(j,k)$, where $j \rmv\in\rmv \{1,\ldots,J\}$ and $k \rmv\in\rmv \Set{K}_n^{(j)} \!\triangleq\rmv \{ 1,\dots,K_n^{(j)} \}$ (which implies that there are $K_n^{(j)}$ PFs for each PA $j$). Note that the set of PFs also includes the PA itself. The number of PAs $J$ is known, but the number of PFs $K_n^{(j)}$ (for PA $j$) is unknown and random. The existence of the $(j,k)$th PF as an actual \emph{feature} is indicated by the variable $r_{k,n}^{(j)} \rmv\!\in\rmv \{0,1\}$, where $r_{k,n}^{(j)} \rmv\!=\rmv 0$ ($r_{k,n}^{(j)} \rmv\!=\! 1$) means that the PF does not exist (exists) at time $n$. The state of PF $(j,k)$ is the PF's position $\V{a}_{k,n}^{(j)}\ist$, and the \emph{augmented state} is $\V{y}_{k,n}^{(j)} \!\triangleq\rmv \big[\V{a}_{k,n}^{(j)\text{T}} \,\ist r_{k,n}^{(j)}\big]\trans$ \cite{MeyerTSP2017}. We also define 
$\V{y}_n^{(j)} \!\triangleq \big[\V{y}_{1,n}^{(j)\text{T}} \rmv\cdots\ist \V{y}_{K_n^{(j)}\rmv,n}^{(j)\text{T}} \big]\trans$ and $\V{y}_n \!\triangleq \big[\V{y}_n^{(1)\text{T}} \rmv\cdots\ist \V{y}_n^{(J)\text{T}} \big]\trans\rmv$. We will formally consider PF states also for the nonexisting PFs (case $r^{(j)}_{k,n} \!\rmv=\rmv 0$); however, their values are obviously irrelevant. Therefore, the probability density function (pdf) of an augmented state, $f\big(\V{y}^{(j)}_{k,n}\big) \rmv=\rmv f\big(\V{a}^{(j)}_{k,n}\ist, r^{(j)}_{k,n}\big)$, is such that for $r^{(j)}_{k,n} \!\rmv=\rmv 0$, $f\big(\V{a}^{(j)}_{k,n}\ist, 0\big) = f^{(j)}_{k,n} \ist f_{\text{D}}\big(\V{a}^{(j)}_{k,n}\big)$, where $f_{\text{D}}\big(\V{a}^{(j)}_{k,n}\big)$ is an arbitrary ``dummy pdf'' and $f^{(j)}_{k,n} \geq 0$ can be interpreted as the probability of PF nonexistence \cite{MeyerTSP2017}. We note that the joint augmented PF state, described here by the random vector $\V{y}_n^{(j)}\rmv$, can also be modeled by a multi-Bernoulli 
random finite set \cite{WilliamsTAES2015}. However, the unordered nature of random finite sets would complicate the development of a factor graph and a BP algorithm. 

At any time $n$, each PF is either a \textit{legacy PF}, which was already established in the past, or a \textit{new PF}, which is established for the first time. The augmented states of legacy PFs and new PFs for PA $j$
will be denoted by $\tilde{\V{y}}_{k,n}^{(j)} \!\triangleq\rmv \big[ \tilde{\V{a}}_{k,n}^{(j)\text{T}} \; \tilde{r}_{k,n}^{(j)} \big]\trans\rmv$, $k \!\in\! \Set{K}_{n-1}^{(j)}$ and $\breve{\V{y}}_{m,n}^{(j)} \!\triangleq\rmv \big[ \breve{\V{a}}_{m,n}^{(j)\text{T}} \; \breve{r}_{m,n}^{(j)} \big]\trans\rmv$, $m \!\in\! \Set{M}_{n}^{(j)}\!$, respectively. Thus, the number of new PFs equals the number of measurements, $M^{(j)}_n \rmv$. The set and number of legacy PFs are updated according to
\begin{equation}
\Set{K}_{n}^{(j)} \rmv= \Set{K}_{n-1}^{(j)} \rmv\cup \Set{M}_{n}^{(j)} , \qquad\; K_{n}^{(j)} \rmv= K_{n-1}^{(j)} + M_{n}^{(j)},
\label{eq:numandsetFeat}
\end{equation}
where the first relation is understood to include a suitable reindexing of the elements of $\Set{M}_{n}^{(j)}\rmv$. (The number of PFs does not actually grow by $M_{n}^{(j)}$ because the set of PFs is pruned, i.e., PFs with small existence probability are discarded, as explained in Section~\ref{sec:problem}.) We also define the following state-related vectors. For the legacy PFs for PA $j$, 
$\tilde{\V{a}}_n^{(j)} \!\triangleq \big[\tilde{\V{a}}_{1,n}^{(j)\text{T}} \rmv\cdots\ist \tilde{\V{a}}_{K_{n-1}^{(j)}\rmv,n}^{(j)\text{T}} \big]\trans\rmv$, 
$\tilde{\V{r}}^{(j)}_n \triangleq \big[\tilde{r}_{1,n}^{(j)} \cdots\ist \tilde{r}_{\!K^{(j)}_{n-1}\!,n}^{(j)}\big]\trans\rmv$, and $\tilde{\V{y}}_n^{(j)} \!\triangleq \big[\tilde{\V{y}}_{1,n}^{(j)\text{T}} \rmv\cdots\ist \tilde{\V{y}}_{K_{n-1}^{(j)}\rmv,n}^{(j)\text{T}} \big]\trans\rmv$. For the new PFs for PA $j$,  $\breve{\V{a}}_n^{(j)} \!\triangleq \big[\breve{\V{a}}_{1,n}^{(j)\text{T}} \rmv\cdots\ist \breve{\V{a}}_{M_{n}^{(j)}\rmv,n}^{(j)\text{T}} \big]\trans\rmv$, $\breve{\V{r}}^{(j)}_n \triangleq \big[\breve{r}_{1,n}^{(j)} \cdots\ist \breve{r}_{\!M^{(j)}_n\!,n}^{(j)}\big]\trans\rmv$,
and $\breve{\V{y}}_n^{(j)} \!\triangleq \big[\breve{\V{y}}_{1,n}^{(j)\text{T}} \rmv\cdots\ist \breve{\V{y}}_{M_{n}^{(j)}\rmv,n}^{(j)\text{T}} \big]\trans\rmv$.
For the combination of legacy PFs and new PFs for PA $j$, $\V{y}_n^{(j)} \!\triangleq \big[\tilde{\V{y}}_n^{(j)\text{T}} \,\ist \breve{\V{y}}_n^{(j)\text{T}}\big]\trans\rmv$;
note that the vector entries (subvectors) of $\V{y}_n^{(j)}$ are given by $\V{y}_{k,n}^{(j)}$ for $k \in \Set{K}_{n}^{(j)} \!= \Set{K}_{n-1}^{(j)} \rmv\cup \Set{M}_{n}^{(j)}$\rmv.
For all the legacy PFs, $\tilde{\V{y}}_n \!\triangleq \big[\tilde{\V{y}}_n^{(1)\text{T}} \rmv\cdots\ist \tilde{\V{y}}_n^{(J)\text{T}} \big]\trans\rmv$, and for all the new PFs, 
$\breve{\V{y}}_n \!\triangleq \big[\breve{\V{y}}_n^{(1)\text{T}} \rmv\cdots\ist \breve{\V{y}}_n^{(J)\text{T}} \big]\trans\rmv$.

The number of new PFs at time $n$ is known only after the current measurements have been observed. Features that are observed for the first time will be referred to as \emph{newly detected features}. Before the current measurements are observed, only prior information about the newly detected features is available (namely, their prior distribution and mean number). After the current measurements are observed, newly detected features are represented by new PFs.

The agent state $\V{x}_n$ and the augmented states of the legacy PFs, $\tilde{\V{y}}_{k,n}^{(j)}$, are assumed to evolve independently according to Markovian state dynamics, i.e.,
\begin{align}
f\big(\V{x}_n,\tilde{\V{y}}_n|\V{x}_{n-1},\V{y}_{n-1}\big) &=  f(\V{x}_{n}|\V{x}_{n-1})f(\tilde{\V{y}}_{n}|\V{y}_{n-1}) \nn\\
&= f(\V{x}_{n}|\V{x}_{n-1})  \nn \\ 
&\hspace{5mm}\times \prod_{j=1}^J\prod_{k=1}^{K_{n-1}^{(j)}} \!  f\big(\tilde{\V{y}}_{k,n}^{(j)} \big| \V{y}_{k, n-1}^{(j)}\big) \ist,
\label{eq:state_space}
\end{align}
where $f(\V{x}_{n}|\V{x}_{n-1})$ and $f\big(\tilde{\V{y}}_{k,n}^{(j)} \big| \V{y}_{k, n-1}^{(j)}\big) = f\big(\tilde{\V{a}}_{k,n}^{(j)},\tilde{r}_{k,n}^{(j)} \big| $\linebreak$\V{a}_{k,n-1}^{(j)}, r_{k,n-1}^{(j)}\big)$ are the state-transition pdfs of the agent and of legacy PF $(j,k)$, respectively. 
If PF $(j,k)$ exists at time $n \rmv-\! 1$, i.e., $r_{k,n-1}^{(j)} \!=\! 1$, it either dies, i.e., $\tilde{r}_{k,n}^{(j)} \!=\rmv 0$, or survives, i.e., $\tilde{r}_{k,n}^{(j)} \!=\! 1$; in the latter case, it becomes a legacy PF at time $n$. The probability of survival is denoted by $P_\text{s}$. If the PF survives, its new state $\tilde{\V{a}}_{k,n}^{(j)}$ is distributed according to the state-transition pdf $f\big(\tilde{\V{a}}_{k,n}^{(j)} \big| \V{a}_{k,n-1}^{(j)}\big)$.
Therefore, $f\big(\tilde{\V{y}}_{k,n}^{(j)} \big| \V{y}_{k, n-1}^{(j)}\big)$ in \eqref{eq:state_space} is given for $r_{k,n-1}^{(j)} \rmv=\rmv 1$ by 
\begin{align}
&f\big(\tilde{\V{a}}_{k,n}^{(j)},\tilde{r}_{k,n}^{(j)} \big| \V{a}_{k,n-1}^{(j)}, r_{k,n-1}^{(j)} \!=\! 1\big) \nn \\[1.5mm]
&\hspace{10mm}= \begin{cases} 
    (1 \!-\rmv P_\text{s}) \ist f_\text{D}\big(\tilde{\V{a}}_{k,n}^{(j)}\big) , &\!\!\! \tilde{r}_{k,n}^{(j)} \!=\rmv 0 \\[1mm]
    P_\text{s} \ist f\big(\tilde{\V{a}}_{k,n}^{(j)} \big| \V{a}_{k,n-1}^{(j)}\big) , &\!\!\! \tilde{r}_{k,n}^{(j)} \!=\! 1.
\end{cases} \!\!\!\!
\label{eq:survival_transition}
\end{align}
If PF $(j,k)$ does not exist at time $n \rmv-\! 1$, i.e., $r_{k,n-1}^{(j)} \!=\! 0$, it cannot exist as a legacy PF at time $n$ either. Therefore, 
\be
\hspace*{-3mm}
f\big(\tilde{\V{a}}_{k,n}^{(j)},\tilde{r}_{k,n}^{(j)} \big| \V{a}_{k,n-1}^{(j)}, r_{k,n-1}^{(j)} \!=\rmv 0\big)
\rmv=\rmv\begin{cases} 
    f_\text{D}\big(\tilde{\V{a}}_{k,n}^{(j)}\big) , &\!\!\! \tilde{r}_{k,n}^{(j)} \!=\rmv 0 \\[1mm]
    0 , &\!\!\! \tilde{r}_{k,n}^{(j)} \!=\! 1.
  \end{cases} \!\!\!\!
  \label{eq:dummy_transition}
\ee

\subsection{Association Vectors}
\label{sec:assoc_vec_description}
For each PA $j$, the measurements (MPC parameter estimates) $\V{z}_{m,n}^{(j)}\ist$, $m \rmv\in\rmv \Set{M}_n^{(j)}$ described in Section~\ref{sec:signalmodel} are subject to a measurement origin uncertainty, also known as DA uncertainty. That is, it is not known which measurement $\V{z}_{m,n}^{(j)}\ist$ is associated with which PF $k \!\in\! \Set{K}_n^{(j)}\rmv$, or if a measurement $\V{z}_{m,n}^{(j)}$ did not originate from any PF (this is known as a \emph{false alarm} or \emph{clutter}), or if a PF did not give rise to any measurement (this is known as a \emph{missed detection}). The probability that a PF is ``detected'' in the sense that it generates a measurement $\V{z}_{m,n}^{(j)}$ in the MPC parameter estimation stage is denoted by $P_\text{d}^{(j)}\big(\V{x}_n ,\V{a}_{k,n}^{(j)}\big)$. The distribution of false alarm measurements is described by the pdf $f_{\text{FA}}\big( \V{z}_{m,n}^{(j)} \big)$. The functions $P_\text{d}^{(j)}\big(\V{x}_n ,\V{a}_{k,n}^{(j)}\big) \!\in\rmv (0,1]$ and $f_{\text{FA}}\
big( \V{z}_{m,n}^{(j)} \big) \!\ge\rmv 0$ are supposed known. Following \cite{BarShalom11}, we assume that at any time $n$, each PF can generate at most one measurement, and each measurement can be generated by at most one PF. 

The associations between the measurements $m \in \Set{M}_{n}^{(j)}$ and the legacy PF states $(j,k)$, $k \!\in\! \Set{K}_{n-1}^{(j)}$ at time $n$ can be described by the $K_{n-1}^{(j)}$-dimensional \textit{feature-oriented DA vector} $\V{c}_n^{(j)} = \big[c_{1,n}^{(j)} \cdots\ist c_{K_{n-1}^{(j)},n}^{(j)} \big]\trans\rmv$, whose $k$th entry is defined to be $c_{k,n}^{(j)} \!\triangleq\rmv m \!\in\! \Set{M}_n^{(j)}$ if legacy PF $(j,k)$  generates measurement $\V{z}_{m,n}^{(j)}$, and $c_{k,n}^{(j)} \!\triangleq\rmv 0$ if it does not generate any measurement. In addition, following \cite{WilliamsTAE2014,MeyerTSP2017}, we consider the $M_n^{(j)}$-dimensional \textit{measurement-oriented DA vector} $\V{b}_{n}^{(j)} = \big[b_{1,n}^{(j)} \cdots\ist b_{M_n^{(j)}\!,n}^{(j)}\big]\trans\rmv$, whose $m$th entry is defined to be  
$b_{m,n}^{(j)} \!\triangleq\rmv k \!\in\! \Set{K}_{n-1}^{(j)}$ if measurement $\V{z}_{m,n}^{(j)}$ is generated by legacy PF $(j,k)$, and $b_{m,n}^{(j)} \!\triangleq\rmv 0$ if it is not generated by any legacy PF. We also define $\V{c}_n \rmv\triangleq\rmv \big[\V{c}_n^{(1)\text{T}} \rmv\cdots\ist \V{c}_n^{(J)\text{T}} \big]\trans\rmv$ and $\V{b}_n \rmv\triangleq\rmv \big[{\V{b}_n^{(1)\text{T}}} \rmv\cdots\ist {\V{b}_n^{(J)\text{T}}} \big]\trans\rmv$. The two DA vectors $\V{c}_n$ and $\V{b}_n$ are unknown and modeled as random. They are equivalent since one can be determined from the other. However, the redundant formulation of DA uncertainty in terms of both $\V{c}_n$ and $\V{b}_n$ is key to obtaining the scalability properties of the BP algorithm to be presented in Section~\ref{sec:BP_algorithm}. Furthermore, as will be discussed in Section \ref{sec:FGderivation}, it facilitates the establishment of a factor graph for the problem of jointly inferring the agent state and the states of the legacy PFs and new 
PFs.

\subsection{Prior Distributions}
\label{eq:JointAssociation}

For a given PA $j$, there are $M^{(j)}_n \!= |\Set{M}^{(j)}_n|$ new PFs at time $n$. The number of false alarms and the number of newly detected features are 
assumed Poisson distributed with mean $\mu^{(j)}_\text{FA}\rmv$ and $\mu^{(j)}_{\text{n},n}$, respectively \cite{BarShalom11,vermaak05}. Then, one can derive the following expression of the joint conditional prior probability mass function (pmf) of the DA vector $\V{c}_{n}^{(j)}\rmv$, the vector of existence indicators of new PFs, $\breve{\V{r}}^{(j)}_n\rmv$, and the number of measurements or equivalently new PFs, $M_{n}^{(j)}\rmv$, given the state of the mobile agent, $\V{x}_n$, and the vector of augmented states of legacy PFs, $\tilde{\V{y}}_{n}^{(j)}$ (cf.\ \cite{MeyerTSP2017})
\begin{align}
&p\big(\V{c}_{n}^{(j)} \rmv, \breve{\V{r}}^{(j)}_n \rmv, M_{n}^{(j)} \big|\V{x}_n , \tilde{\V{y}}_{n}^{(j)}\big) \nn \\
   &=\ist \chi_{\V{c}_{n}^{(j)} \!,\ist \breve{\V{r}}^{(j)}_n \!,\ist M_{n}^{(j)} } \, \Psi\big(\V{c}_{n}^{(j)}\big) 
     \Bigg(\prod_{m \in \Set{N}_{\!\breve{\V{r}}^{(j)}_{n}}} \!\!\!\! \Gamma_{\!\V{c}_{n}^{(j)}\rmv}\big(\breve{r}_{m,n}^{(j)}\big) \!\Bigg) \nn \\
   &\hspace{6mm} \times \Bigg( \prod_{k \in \Set{D}_{\!\V{c}_{n}^{(j)}\!\rmv, \tilde{\V{r}}_{n}^{(j)}}} \!\!\!\!\! P_{\text{d}}^{(j)}\big(\V{x}_n\rmv ,\V{a}_{k,n}^{(j)}\big) \! \Bigg) \nn\\[-.5mm]
&\hspace{6mm} \times \!\prod_{k'\rmv \in \bar{\Set{D}}_{\!\V{c}_{n}^{(j)}\!\rmv, \tilde{\V{r}}_{n}^{(j)}}} \!\!\!\!\big[1\big(c_{k'\!,n}^{(j)}\big) - \tilde{r}_{k'\!,n}^{(j)} 
  P_{\text{d}}^{(j)}\big(\V{x}_n\rmv,\tilde{\V{a}}_{k'\!,n}^{(j)}\big) \big] \ist,
\label{eq:assocpriorSens}\\[-7.8mm]
\nn
\end{align}
with $\chi_{\V{c}_{n}^{(j)} \!,\ist \breve{\V{r}}^{(j)}_n \!,\ist M_{n}^{(j)} } \!\triangleq\rmv e^{-(\mu^{(j)}_\text{FA}+\mu_{\text{n},n}^{(j)})}\ist(\mu_{\text{n},n}^{(j)})^{\big|\Set{N}_{\!\breve{\V{r}}_{n}^{(j)}}\big|} $ $\times  (\mu^{(j)}_\text{FA})^{M_{n}^{(j)} - \big|\Set{D}_{\!\V{c}_{n}^{(j)}\!\rmv, \tilde{\V{r}}_{n}^{(j)}}\big| - \big|\Set{N}_{\!\breve{\V{r}}_{n}^{(j)}}\big|} /\ist M_{n}^{(j)}!$. Here $\Set{N}_{\rmv\breve{\V{r}}^{(j)}_{n}}\!$ denotes the set of existing new PFs, i.e., $\Set{N}_{\rmv\breve{\V{r}}^{(j)}_{n}} \!\triangleq \big\{m \rmv\in\rmv \Set{M}_n^{(j)} \!:\breve{r}_{m,n}^{(j)} \!=\! 1 \big\}$;
$\Set{D}_{\!\V{c}_{n}^{(j)}\!\rmv, \tilde{\V{r}}_{n}^{(j)}}$ denotes the set of existing legacy PFs for PA $j$, i.e., 
$\Set{D}_{\!\V{c}_{n}^{(j)}\!\rmv, \tilde{\V{r}}_{n}^{(j)}} \!\triangleq \big\{k \rmv\in\rmv \Set{K}_{n-1}^{(j)} \!: \tilde{r}_{k,n}^{(j)} \!=\! 1, c_{k,n}^{(j)} \!\neq\rmv 0 \big\}$; 
and $\Set{\bar{D}}_{\!\V{c}_{n}^{(j)}\!\rmv, \tilde{\V{r}}_{n}^{(j)}}$
$\triangleq \Set{K}_{n-1}^{(j)} \rmv\big\backslash \Set{D}_{\!\V{c}_{n}^{(j)}\!\rmv, \tilde{\V{r}}_{n}^{(j)}}$. Furthermore,
$\Psi\big(\V{c}_{n}^{(j)}\big)$ is defined to be $0$ if there exist $k,k' \!\in\rmv \Set{K}_{n-1}^{(j)}$ with $k \!\neq\! k'\rmv$ such that $c_{k,n}^{(j)} \!=\rmv c_{k',n}^{(j)} \!\neq\rmv 0$,
and to be $1$ otherwise; and $\Gamma_{\!\V{c}_{n}^{(j)}\rmv}\big(\breve{r}_{m,n}^{(j)}\big)$ is defined to be $0$ if $\breve{r}_{m,n}^{(j)} \!=\! 1$ and there exists 
$k \!\in\rmv \Set{K}_{n-1}^{(j)}$ such that $c_{k,n}^{(j)} \!\rmv=\rmv m$, and to be $1$ otherwise. Finally, $1(c)$ denotes the indicator function of the event $c \rmv=\rmv 0$ (i.e., $1(c) \rmv=\rmv 1$ if $c \rmv=\rmv 0$ and $0$ otherwise). The functions $\Psi\big(\V{c}_{n}^{(j)}\big)$ and $\Gamma_{\V{c}_{n}^{(j)}}\big(\breve{r}_{m,n}^{(j)}\big)$ enforce our DA assumption from Section \ref{sec:assoc_vec_description}, i.e.,
$\Psi\big(\V{c}_{n}^{(j)}\big)$ enforces $p\big(\V{c}_{n}^{(j)} \rmv, \breve{\V{r}}^{(j)}_n\rmv, M_{n}^{(j)} \big| \V{x}_n \rmv, \tilde{\V{y}}_{n}^{(j)} \big) \rmv=\rmv 0$ if any measurement is associated with more than one legacy PF, and $\Gamma_{\V{c}_{n}^{(j)}}\big(\breve{r}_{m,n}^{(j)}\big)$ enforces 
$p\big(\V{c}_{n}^{(j)} \rmv, \breve{\V{r}}^{(j)}_n\rmv, M_{n}^{(j)} \big|\V{x}_n \rmv, \tilde{\V{y}}_{n}^{(j)}\big) \rmv=\rmv 0$ if a new PF is associated with a measurement $m$ that is also associated with a legacy PF. At time $n \rmv=\! 1$, prior information on PAs and VAs can be incorporated by introducing legacy PFs. Alternatively, there are no legacy PFs at $n \rmv=\! 1$, i.e., $\tilde{\V{y}}^{(j)}_1\rmv$ is an empty vector, and thus $p\big(\V{c}^{(j)}_{1} \rmv, \breve{\V{r}}^{(j)}_{1} \rmv,  M^{(j)}_{1} \big| \V{x}_1 \rmv, \tilde{\V{y}}_{1}^{(j)}\big) \rmv= p\big(\V{c}^{(j)}_{1}, \breve{\V{r}}^{(j)}_{1} \rmv,M^{(j)}_{1}\big|\V{x}_1\big)$. 
An expression of this pmf can be obtained by replacing in \eqref{eq:assocpriorSens} (for $n \rmv=\! 1$) all factors involving 
$\tilde{\V{y}}^{(j)}_{1}$ (or, equivalently, $\tilde{\V{a}}_1^{(j)}$ and $\tilde{\V{r}}^{(j)}_1$) by $1$. 

The states of newly detected features are assumed a priori independent and identically distributed (iid) according to some pdf  $f_{\text{n},n}\big(\breve{\V{a}}^{(j)}_{m,n}\big| \V{x}_n\big)$, to be discussed in Section \ref{sec:sim_param}. The prior pdf of the states of new PFs for PA $j$, $\breve{\V{a}}_n^{(j)}\rmv$, conditioned on $\V{x}_n$, $\breve{\V{r}}^{(j)}_{n}\rmv$, and $M^{(j)}_{n}\rmv$ is then obtained as
\begin{align}
\! f\big(\breve{\V{a}}^{(j)}_{n} \big|\V{x}_n , \breve{\V{r}}^{(j)}_{n}\rmv , M^{(j)}_{n} \big) 
&= \Bigg( \prod_{m \in \Set{N}_{\!\breve{\V{r}}^{(j)}_{n}}} \!\!\!\! f_{\text{n},n}\big(\breve{\V{a}}^{(j)}_{m,n}\big |\V{x}_n\big) \rmv\Bigg) \nn \\
&\hspace{6mm} \times \prod_{m' \in \bar{\Set{N}}_{\!\breve{\V{r}}^{(j)}_{n}}} \!\!\!\!\rmv f_{\text{D}}\big(\breve{\V{a}}^{(j)}_{m',n}\big) \ist,
\label{eq:factorization_trans2}
\end{align}
where $\bar{\Set{N}}_{\rmv\breve{\V{r}}^{(j)}_{n}} \!\rmv\triangleq\rmv \Set{M}_{n}^{(j)} \ist\big\backslash\, \Set{N}_{\rmv\breve{\V{r}}^{(j)}_{n}}$. Note that before the measurements are obtained, $M^{(j)}_{n}$ and, thus, the length of the vectors $\breve{\V{a}}^{(j)}_{n}$ and $\breve{\V{r}}^{(j)}_{n}$ (which is $M^{(j)}_{n}$) is random.

We assume that for the agent state at time $n \rmv=\! 1$, $\V{x}_{1}$, an informative prior pdf $f(\V{x}_{1})$ is available. 
We also assume that the new PF state vector $\breve{\V{a}}^{(j)}_{n}$ and the DA vector $\V{c}^{(j)}_{n}$ are conditionally independent given the legacy PF state vector $\tilde{\V{a}}^{(j)}_{n}\rmv$. Let us next consider $\V{x}_{n'}$, $\V{y}_{n'}$, $\V{c}_{n'}$, and 
$\V{m}_{n'} \rmv\triangleq \big[M^{(1)}_{n'} \cdots\ist M^{(J)}_{n'}\big]\trans$ for all time steps $n' \!=\rmv 1,\dots,n$, and accordingly define 
the vector $\V{x}_{1:n} \!\triangleq\! \big[\V{x}_{1}^{\text{T}} \cdots\ist \V{x}_n^{\text{T}} \big]\trans$ and similarly $\V{y}_{1:n}\ist$, $\V{c}_{1:n}\ist$, and $\V{m}_{1:n}\ist$.
We then obtain the joint prior pdf as
\begin{align}
&f\big( \V{x}_{1:n}, \V{y}_{1:n}, \V{c}_{1:n}, \V{m}_{1:n} \big) \nn \\
&= f(\V{x}_{1}) \Bigg( \prod^{J}_{j'=1} \rmv f\big(\breve{\V{y}}^{(j')}_{1} \rmv, \V{c}^{(j')}_{1} \rmv,  M^{(j')}_{1} \big|\V{x}_1 \big) \!\Bigg) \prod^{n}_{n'=2} \!\rmv f(\V{x}_{n'}|\V{x}_{n'-1} ) \nn\\[.3mm]
&\hspace{5mm}\times \prod^{J}_{j=1}\Bigg(\prod_{k=1}^{K_{n-1}^{(j)}}\rmv\rmv\rmv f\big(\tilde{\V{y}}^{(j)}_{k,n'} \big| \V{y}^{(j)}_{k,n'-1}\big)\Bigg) f\big(\breve{\V{a}}^{(j)}_{n'} \big| \V{x}_{n'} , \breve{\V{r}}^{(j)}_{n'}\rmv , M^{(j)}_{n'}\big) \nn\\[.5mm]
&\hspace{5mm}\times p\big(\V{c}^{(j)}_{n'} \rmv, \breve{\V{r}}^{(j)}_{n'} \rmv,  M^{(j)}_{n'} \big|\V{x}_{n'}, \tilde{\V{y}}^{(j)}_{n'}\big) \ist,
\label{eq:allTimesPrior}
\end{align}
where $f\big(\breve{\V{y}}^{(j)}_{1} \rmv, \V{c}^{(j)}_{1} \rmv,  M^{(j)}_{1}\big|\V{x}_1  \big) =  p\big(\V{c}^{(j)}_{1} \rmv, \breve{\V{r}}^{(j)}_{1} \rmv,  M^{(j)}_{1}\big|\V{x}_1 \big)$ $\times f\big(\breve{\V{a}}^{(j)}_{1} \big| \V{x}_{1} , \breve{\V{r}}^{(j)}_{1}\rmv , M^{(j)}_{1}\big)$ (see \eqref{eq:assocpriorSens} and \eqref{eq:factorization_trans2}), and where the last three factors in \eqref{eq:allTimesPrior} are given by \eqref{eq:survival_transition} and \eqref{eq:dummy_transition}; \eqref{eq:factorization_trans2}; and \eqref{eq:assocpriorSens}, respectively.

\subsection{Likelihood Function}
\label{sec:meas_likelihood} 

The conditional pdf $f\big(\V{z}^{(j)}_{m,n} \big| \V{x}_n, \V{a}^{(j)}_{k,n} \big)$ characterizing the statistical relation between the measurements $\V{z}^{(j)}_{m,n}$ and the states $\V{x}_n$ and $\V{a}^{(j)}_{k,n}$ depends on the concrete measurement model; an example will be considered in Section~\ref{sec:MeasModel}. 
This pdf is a central element in the conditional pdf of the total measurement vector $\V{z}_n \rmv\triangleq\rmv \big[\V{z}_n^{(1)\text{T}}\ist \cdots\ist \V{z}_n^{(J)\text{T}} \big]\trans\rmv$
given $\V{x}_n$, $\tilde{\V{y}}_n$, $\breve{\V{y}}_n$, $\V{c}_n$, and $\V{m}_n$. Assuming that the $\V{z}_{n}^{(j)}$ are conditionally independent across $j$ given $\V{x}_n$, $\tilde{\V{y}}_n$, $\breve{\V{y}}_n$, $\V{c}_n$, and $\V{m}_n$ \cite{BarShalom11}, we obtain $f( \V{z}_n | \V{x}_n,\tilde{\V{y}}_n, \breve{\V{y}}_n, \V{c}_n, \V{m}_n) 
= \prod^{J}_{j = 1} f\big( \V{z}^{(j)}_{n} \big| \V{x}_n, \tilde{\V{y}}^{(j)}_{n} \rmv, \breve{\V{y}}^{(j)}_{n} \rmv, \V{c}^{(j)}_{n} \rmv, M^{(j)}_{n}\big)$, where \cite{BarShalom11}
\begin{align}
&f\big( \V{z}^{(j)}_{n} \big| \V{x}_{n}, \tilde{\V{y}}^{(j)}_{n}\rmv, \breve{\V{y}}^{(j)}_{n}\rmv, \V{c}^{(j)}_{n} \rmv, M^{(j)}_{n}\big) \nn \\
&\hspace{5mm}= \Bigg( \prod^{M^{(j)}_n}_{m = 1} \rmv\rmv f_{\text{FA}}\big( \V{z}^{(j)}_{m,n} \big) \rmv\Bigg) \!\left(\prod_{k\in \Set{D}_{\!\V{c}_{n}^{(j)}\!\rmv, \tilde{\V{r}}_{n}^{(j)}}}\!\!\!\rmv \frac{f\Big( \V{z}^{(j)}_{c^{(j)}_{k,n},n} \Big|\ist \V{x}_n, \tilde{\V{a}}^{(j)}_{k,n} \Big)} {f_{\text{FA}}\Big( \V{z}^{(j)}_{c^{(j)}_{k,n},n} \Big)}\right) \nn \\[.5mm]
	& \hspace{10mm}\times\!\!\prod_{m'\in \Set{N}_{\!\breve{\V{r}}^{(j)}_{n}}}\!\!\!\rmv 
	\frac{f\big( \V{z}^{(j)}_{m'\!,n} \big|\ist \V{x}_n, \breve{\V{a}}^{(j)}_{m'\!,n} \big)}{f_{\text{FA}}\big( \V{z}^{(j)}_{m'\!,n} \big)} \,.
\label{eq:factorization_like_panchor}\\[-6mm]\nn
\end{align}
In particular, at $n \rmv=\! 1$, $f\big( \V{z}^{(j)}_1\big| \V{x}_1,\tilde{\V{y}}^{(j)}_1\rmv,\breve{\V{y}}^{(j)}_1\rmv, \V{c}^{(j)}_1\rmv, M^{(j)}_1\big) 
\rmv=f\big( \V{z}^{(j)}_1\big| \V{x}_1,\breve{\V{y}}^{(j)}_1\rmv, \V{c}^{(j)}_1\rmv, M^{(j)}_1\big)$. An expression of this pdf can be obtained by replacing 
in \eqref{eq:factorization_like_panchor} (for $n \rmv=\! 1$) all factors involving $\tilde{\V{y}}^{(j)}_{1}$ (or, equivalently, $\tilde{\V{a}}_1^{(j)}$ and $\tilde{\V{r}}^{(j)}_1$) by $1$. 

Let us consider $f\big( \V{z}^{(j)}_n\big|\V{x}_n, \tilde{\V{y}}^{(j)}_n\rmv, \breve{\V{y}}^{(j)}_n\rmv, \V{c}^{(j)}_n\rmv, M^{(j)}_{n}\big)$ as a \emph{likelihood function}, 
i.e., a function of $\V{x}_n$, $\tilde{\V{y}}^{(j)}_n\rmv$, $\breve{\V{y}}^{(j)}_n\rmv$, $\V{c}^{(j)}_n\rmv$, and $M^{(j)}_{n}$, for observed $\V{z}^{(j)}_{n}\rmv$. If $\V{z}^{(j)}_{n}$ is observed and therefore fixed, also $ M^{(j)}_{n}$ is fixed, and we can rewrite \eqref{eq:factorization_like_panchor}, 
up to a constant factor, as
\begin{align}
\nn\\[-8mm]
&f\big( \V{z}^{(j)}_n\big|\V{x}_n, \tilde{\V{y}}^{(j)}_n\rmv, \breve{\V{y}}^{(j)}_n\rmv, \V{c}^{(j)}_n\rmv, M^{(j)}_{n}\big) \nn \\ 
&\hspace{10mm}\propto\Bigg( \prod^{K^{(j)}_{n}}_{k=1}\rmv g_1\big( \V{x}_{n}, \tilde{\V{a}}^{(j)}_{k,n}, \tilde{r}^{(j)}_{k,n}, c^{(j)}_{k,n}; \V{z}^{(j)}_{n} \big) \!\Bigg) \nn \\
& \hspace{15mm}\times\!\! \prod_{m\in \Set{N}_{\!\breve{\V{r}}^{(j)}_{n}}}\!\!\!\rmv \frac{f\big( \V{z}^{(j)}_{m,n} \rmv\big|\ist \V{x}_n, \breve{\V{a}}^{(j)}_{m,n} \big)}{f_{\text{FA}}\big( \V{z}^{(j)}_{m,n} \big)} \ist.
\label{eq:factorization_like_panchor2}
\end{align}
Here, 
$g_1\big( \V{x}_{n}, \tilde{\V{a}}^{(j)}_{k,n}, \tilde{r}^{(j)}_{k,n}, c^{(j)}_{k,n}; \V{z}^{(j)}_{n} \big)$ is defined as 
\begin{align}
&g_1\big( \V{x}_{n}, \tilde{\V{a}}^{(j)}_{k,n}, 1, c^{(j)}_{k,n}; \V{z}^{(j)}_{n} \big)\nn\\[1.5mm]
&\hspace{10mm}\triangleq\rmv\rmv \begin{cases}
  \rmv\displaystyle \frac{f\big(\V{z}^{(j)}_{m,n} \big|\V{x}_n,\tilde{\V{a}}^{(j)}_{k,n} \big)}{f_{\text{FA}}\big( \V{z}^{(j)}_{m,n} \big)} \ist, 
  & \! c^{(j)}_{k,n}\!=\rmv m \in\! \Set{M}_n^{(j)} \\[-.5mm]
  \rmv 1 \ist, & \! c^{(j)}_{k,n} \!=\rmv 0
\end{cases}\label{eq:factor1_like} 
\end{align}
and $g_1\big( \V{x}_{n}, \tilde{\V{a}}^{(j)}_{k,n}, 0, c^{(j)}_{k,n}; \V{z}^{(j)}_{n} \big) \rmv\triangleq\rmv 1$. Finally, the likelihood function for $\V{z}_{1:n} \!\triangleq\rmv \big[\V{z}_1\trans \cdots\ist \V{z}_n\trans \big]\trans\rmv$, involving the measurements $\V{z}^{(j)}_{m,n'}$ of all PAs $j = 1,\dots,J$ and all time steps $n' \!=\! 1,\ldots,n$, can be derived similarly to \eqref{eq:allTimesPrior}; one obtains
\begin{align}
\nn\\[-5mm]
&f( \V{z}_{1:n} | \V{x}_{1:n},\rmv \V{y}_{1:n},\rmv \V{c}_{1:n}, \rmv\V{m}_{1:n}) \nn \\[1.5mm]
&\hspace{5mm}\propto\rmv \prod^{J}_{j = 1} \rmv \Bigg( \prod_{m\in \Set{N}_{\!\breve{\V{r}}^{(j)}_{1}}}\!\!\!\rmv \frac{f\big( \V{z}^{(j)}_{m,1} \big|\ist \V{x}_{1} \rmv, \breve{\V{a}}^{(j)}_{m,1} \big)} {f_{\text{FA}}\big( \V{z}^{(j)}_{m,1} \big)} \Bigg) \nn \\[-1mm]
&\hspace{10mm}\times\rmv\rmv\prod^{n}_{n'=2} \rmv\rmv \Bigg(\! \prod_{m'\in \Set{N}_{\!\breve{\V{r}}^{(j)}_{n}}}\!\!\!\rmv \frac{f\big( \V{z}^{(j)}_{m'\!,n'} \big|\ist \V{x}_{n'} , \breve{\V{a}}^{(j)}_{m'\!,n'} \big)}
  {f_{\text{FA}}\big( \V{z}^{(j)}_{m'\!,n'} \big)}  \!\Bigg)\nn\\[-1.5mm]
&\hspace{10mm} \times\prod^{K^{(j)}_{n'}}_{k=1} g_1\big( \V{x}_{n'}, \tilde{\V{a}}^{(j)}_{k,n'}\ist, \tilde{r}^{(j)}_{k,n'}\ist, c^{(j)}_{k,n'}\ist; \V{z}^{(j)}_{n'} \big)\ist. 
\label{eq:allTimeslhf} 
\end{align}

\section{Joint Posterior pdf and Factor Graph}\label{sec:FGderivation}
\subsection{Redundant Formulation of the Exclusion Constraint}
\label{sec:modPrior}

The proposed BP-SLAM algorithm relies on a redundant formulation of probabilistic DA in terms of $\V{c}^{(j)}_{n}$ and $\V{b}^{(j)}_{n}$ \cite{MeyerTSP2017, WilliamsTAE2014}. 
To obtain a probabilistic description and, in turn, a factor graph, we formally replace the exclusion constraint factor $\Psi\big(\V{c}_{n}^{(j)}\big)$ involved in the prior pmf in \eqref{eq:assocpriorSens} by 
\begin{align}
\nn\\[-6mm]
\Psi\big(\V{c}^{(j)}_{n} \rmv,\V{b}^{(j)}_{n}\big) \ist\triangleq\ist \prod^{K^{(j)}_n}_{k=1} \prod^{M^{(j)}_n}_{m=1} \!\psi\big(c^{(j)}_{k,n}\ist, b^{(j)}_{m,n}\big) \ist,\nn
\\[-6mm]\nn
\end{align}
where $\psi\big(c^{(j)}_{k,n}\ist, b^{(j)}_{m,n}\big)$ is defined to be $0$ if either $c^{(j)}_{k,n}\!\rmv=\rmv m$ and $b^{(j)}_{m,n} \!\neq\rmv k$ or 
$b^{(j)}_{m,n}\!=\rmv k$ and $c^{(j)}_{k,n} \!\rmv\neq\rmv m$, and $1$ otherwise. The resulting modified prior pmf $p\big(\V{c}^{(j)}_{n} \rmv,  \V{b}^{(j)}_{n} \rmv, \breve{\V{r}}^{(j)}_{n} \rmv, M^{(j)}_{n} \big| \V{x}_n, \tilde{\V{y}}^{(j)}_{n}\big)$ is related to the original prior pmf $p\big(\V{c}_{n}^{(j)} \rmv, \breve{\V{r}}^{(j)}_n \rmv, M_{n}^{(j)} \big|\V{x}_n, \tilde{\V{y}}_{n}^{(j)}\big)$ in \eqref{eq:assocpriorSens} acording to 
$p\big(\V{c}^{(j)}_{n} \rmv, \breve{\V{r}}^{(j)}_{n} \rmv, M^{(j)}_{n} \big|\V{x}_n, \tilde{\V{y}}^{(j)}_{n} \big) 
= \sum_{\V{b}^{(j)}_{n}} p\big(\V{c}^{(j)}_{n} \rmv, \V{b}^{(j)}_{n} \rmv,\breve{\V{r}}^{(j)}_{n} \rmv, M^{(j)}_{n} \big| \V{x}_n, \tilde{\V{y}}^{(j)}_{n} \big)$. 
Here, the summation is over all $\V{b}^{(j)}_{n} \!\in\rmv \big\{0,1,\ldots, K_n^{(j)}\big\}^{M^{(j)}_n}\!\rmv$, 
where $\{\bigcdot\}^{M_n^{(j)}} \!\!=\rmv \{\bigcdot\} \rmv $\linebreak
$\times\rmv \{\bigcdot\}\times\rmv \ldots \times\rmv \{\bigcdot\}$ denotes the $M^{(j)}_n$-fold Cartesian product.

Let us now consider the product of the likelihood function 
$f\big( \V{z}^{(j)}_n\big|\V{x}_n, \tilde{\V{y}}^{(j)}_n\rmv, \breve{\V{y}}^{(j)}_n\rmv, \V{c}^{(j)}_n\rmv, M^{(j)}_{n}\big)$ in \eqref{eq:factorization_like_panchor2}, 
the prior pdf of the new PF states $f\big(\breve{\V{a}}^{(j)}_{n} \big|\V{x}_n,  \breve{\V{r}}^{(j)}_{n}\rmv, M^{(j)}_{n} \big) $ in \eqref{eq:factorization_trans2}, 
and the modified prior pmf $p\big(\V{c}^{(j)}_{n} \rmv, \V{b}^{(j)}_{n} \rmv, \breve{\V{r}}^{(j)}_{n} \rmv, M^{(j)}_{n} \big|\V{x}_n, \tilde{\V{y}}^{(j)}_{n}\big)$ (cf.\ \eqref{eq:assocpriorSens}). We obtain
\begin{align}
&f\big( \V{z}^{(j)}_{n} \big| \V{x}_n, \tilde{\V{y}}^{(j)}_{n}\rmv,\breve{\V{y}}^{(j)}_{n}, \V{c}^{(j)}_{n} \rmv, M^{(j)}_{n}\big) \ist f\big(\breve{\V{a}}^{(j)}_{n}\big |\V{x}_n, \breve{\V{r}}^{(j)}_{n}\rmv, M^{(j)}_{n}) \nn \\[1mm]
&\hspace{5mm}\times p\big(\V{c}^{(j)}_{n} \rmv, \V{b}^{(j)}_{n} \rmv, \breve{\V{r}}^{(j)}_{n} \rmv, M^{(j)}_{n} \big|\V{x}_n, \tilde{\V{y}}^{(j)}_{n} \big) \nn\\[-.5mm]
&\propto \Psi\big(\V{c}^{(j)}_{n},\V{b}^{(j)}_{n}\big) \Bigg(\rmv\prod^{K^{(j)}_n}_{k=1} \rmv g\big( \V{x}_n, \tilde{\V{a}}^{(j)}_{k,n} , \tilde{r}^{(j)}_{k,n}, c^{(j)}_{k,n}; \V{z}^{(j)}_{n} \big) \!\Bigg) \! \nn\\[0mm]
&\hspace{0mm}\times \Bigg(\prod_{m \in \Set{N}_{\!\breve{\V{r}}^{(j)}_{n}}} \!\!\!\!\rmv\rmv \frac{\mu_{\text{n}, n}^{(j)} \rmv f_{\text{n},n}\big(\breve{\V{a}}^{(j)}_{m,n}\rmv\big| \V{x}_n\big) \ist \Gamma_{\!\V{c}_{n}^{(j)}\rmv}\big(\breve{r}_{m,n}^{(j)}\big)\ist f\big( \V{z}^{(j)}_{m,n} \rmv\big|\ist \V{x}_n, \breve{\V{a}}^{(j)}_{m,n} \big)}{f_{\text{FA}}\big( \V{z}^{(j)}_{m,n} \big)}\Bigg) \nn \\[-1mm]
&\hspace{0mm}\times \Bigg(\prod_{m' \in \bar{\Set{N}}_{\!\breve{\V{r}}^{(j)}_{n}}}f_{\text{D}}\big(\breve{\V{a}}^{(j)}_{m'\!,n}\big) \! \Bigg)\ist.\label{eq:factorGraph_0}\\[-8mm]\nn
\end{align}
Here, $g\big( \V{x}_n, \tilde{\V{a}}^{(j)}_{k,n} , \tilde{r}^{(j)}_{k,n}, c^{(j)}_{k,n}; \V{z}^{(j)}_{n} \big) \rmv\rmv \triangleq \rmv\rmv g_1\big( \V{x}_n, \tilde{\V{a}}^{(j)}_{k,n} , \tilde{r}^{(j)}_{k,n},  c^{(j)}_{k,n}; $\linebreak$\V{z}^{(j)}_{n}\big)g_2\big( \V{x}_n, \tilde{\V{a}}^{(j)}_{k,n} , \tilde{r}^{(j)}_{k,n}, c^{(j)}_{k,n}; M^{(j)}_{n} \big)$, where $g_1\big( \V{x}_n, \tilde{\V{a}}^{(j)}_{k,n} , \tilde{r}^{(j)}_{k,n}, $\linebreak$ c^{(j)}_{k,n}; \V{z}^{(j)}_{n} \big)$ was defined in and below \eqref{eq:factor1_like} and $g_2\big( \V{x}_n, \tilde{\V{a}}^{(j)}_{k,n},$\linebreak$ \tilde{r}^{(j)}_{k,n}, c^{(j)}_{k,n}; M^{(j)}_{n} \big)$ is defined as
\begin{align}
&g_2\big( \V{x}_n, \tilde{\V{a}}^{(j)}_{k,n}, 1, c^{(j)}_{k,n}; M^{(j)}_{n} \big) \nn \\[1.5mm]
&\hspace{10mm}\triangleq \begin{cases}
    \displaystyle \frac{P^{(j)}_\text{d}\big(\V{x}_n ,\tilde{\V{a}}^{(j)}_{k,n}\big)}{\mu^{(j)}_\text{FA}} \ist, 
      &\rmv\rmv c^{(j)}_{k,n}\! \in\! \Set{M}^{(j)}_n\\[4.5mm]
     1 \!-\rmv P^{(j)}_\text{d}\big( \V{x}_n ,\tilde{\V{a}}^{(j)}_{k,n} \big) \ist, & \rmv\rmv c^{(j)}_{k,n} \!=\rmv 0
  \end{cases} \nn
\end{align}
and $g_2\big( \V{x}_n, \tilde{\V{a}}^{(j)}_{k,n}, 0, c^{(j)}_{k,n}; M^{(j)}_{n} \big) \rmv\triangleq 1\big(c^{(j)}_{k,n}\big)$. 
One thus obtains for 
$g\big( \V{x}_n, \tilde{\V{a}}^{(j)}_{k,n} , \tilde{r}^{(j)}_{k,n}, c^{(j)}_{k,n}; \V{z}^{(j)}_{n} \big)$
\begin{align}
&g\big( \V{x}_n, \tilde{\V{a}}^{(j)}_{k,n} , 1, c^{(j)}_{k,n}; \V{z}^{(j)}_{n} \big)  \nn \\[1.5mm]
&\hspace{3mm}=\begin{cases}
    \displaystyle \frac{P_{\text{d}}^{(j)}\big(\V{x}_n ,\tilde{\V{a}}^{(j)}_{k,n}\big) \ist f\big( \V{z}^{(j)}_{m,n} \big| \V{x}_n, \tilde{\V{a}}^{(j)}_{k,n} \big)}{\mu^{(j)}_\text{FA} \ist f_{\text{FA}}\big( \V{z}^{(j)}_{m,n} \big)} \ist, 
      &\rmv c^{(j)}_{k,n}\!=\rmv m \in\! \Set{M}^{(j)}_n \nn\\[4.5mm]
     1 \!-\rmv P_{\text{d}}^{(j)}\big(\V{x}_n ,\tilde{\V{a}}^{(j)}_{k,n}\big) \ist, &\rmv c^{(j)}_{k,n} \!=\rmv 0 
  \end{cases}\nn\\[-10mm]\nn
\end{align}
and $g\big( \V{x}_n, \tilde{\V{a}}^{(j)}_{k,n} , 0, c^{(j)}_{k,n}; \V{z}^{(j)}_{n} \big) = 1\big(c^{(j)}_{k,n}\big)$. For a choice of $\V{c}^{(j)}_{n}$ and $\V{b}^{(j)}_{n}$ that is valid in the sense that $p\big(\V{c}^{(j)}_{n} \rmv, \V{b}^{(j)}_{n} \rmv, \breve{\V{r}}^{(j)}_{n} \rmv, M^{(j)}_{n} \big|\V{x}_n, \tilde{\V{y}}^{(j)}_{n}\big) \rmv\neq\rmv 0$, the exclusion constraint expressed by $\Gamma_{\!\V{c}_{n}^{(j)}\rmv}\big(\breve{r}_{m,n}^{(j)}\big)$ is satisfied, i.e., $\Gamma_{\!\V{c}_{n}^{(j)}\rmv}\big(\breve{r}_{m,n}^{(j)}\big) \rmv=\! 1$, if and only if both $b^{(j)}_{m,n} \!=\rmv 0$ and $\breve{r}^{(j)}_{m,n} \!=\! 1$. Therefore, in \eqref{eq:factorGraph_0}, we can summarize the products over all $m \!\in\! \Set{N}_{\rmv\breve{\V{r}}^{(j)}_{n}}\!$ and over all  $m' \! \in\! \bar{\Set{N}}_{\rmv\breve{\V{r}}^{(j)}_{n}}\!$ by a product over all $m \rmv\in\rmv \Set{M}^{(j)}_n=\rmv\{1,\ldots, M^{(j)}_n\}$. More specifically, we can rewrite \eqref{eq:factorGraph_0} as 
\begin{align}
&f\big( \V{z}^{(j)}_{n} \big| \V{x}_n, \tilde{\V{y}}^{(j)}_{n}\rmv,\breve{\V{y}}^{(j)}_{n}, \V{c}^{(j)}_{n} \rmv, M^{(j)}_{n}\big) \ist f\big(\breve{\V{a}}^{(j)}_{n}\big | \V{x}_n, \breve{\V{r}}^{(j)}_{n}\rmv, M^{(j)}_{n}) \nn \\[1mm]
&\hspace{5mm} \times p\big(\V{c}^{(j)}_{n} \rmv, \V{b}^{(j)}_{n} \rmv, \breve{\V{r}}^{(j)}_{n} \rmv, M^{(j)}_{n} \big|\V{x}_n, \tilde{\V{y}}^{(j)}_{n} \big) \nn\\[-.5mm]
&\hspace{2mm}\propto \Psi\big(\V{c}^{(j)}_{n},\V{b}^{(j)}_{n}\big) \Bigg(\rmv\prod^{K^{(j)}_{n-1}}_{k=1} \rmv g\big( \V{x}_n, \tilde{\V{a}}^{(j)}_{k,n} , \tilde{r}^{(j)}_{k,n}, c^{(j)}_{k,n}; \V{z}^{(j)}_{n} \big)\!\Bigg)\nn\\[-1.5mm]
&\hspace{7mm}\times \prod^{M^{(j)}_n}_{m=1} \! h\big( \V{x}_n, \breve{\V{a}}^{(j)}_{m,n} , \breve{r}^{(j)}_{m,n}, b^{(j)}_{m,n}; \V{z}^{(j)}_{n} \big) \ist,
\label{eq:factorGraph} 
\end{align}
where $h\big( \V{x}_n, \breve{\V{a}}^{(j)}_{m,n} , \breve{r}^{(j)}_{m,n}, b^{(j)}_{m,n}; \V{z}^{(j)}_{n} \big)$ is defined as
\begin{align}
&h\big( \V{x}_n, \breve{\V{a}}^{(j)}_{m,n} , 1, b^{(j)}_{m,n}; \V{z}^{(j)}_{n} \big) \nn \\[1mm]
&\triangleq \begin{cases}
0 \ist,  &\rmv\rmv\rmv\rmv\rmv\rmv b^{(j)}_{m,n} \!\in\rmv\Set{K}^{(j)}_{n-1} \\[1.5mm]
\displaystyle \frac{\mu_{\text{n}, n}^{(j)} \ist f_{\text{n},n}\big(\breve{\V{a}}^{(j)}_{m,n}\big| \V{x}_n\big)\rmv\rmv f\big( \V{z}^{(j)}_{m,n} \big| \V{x}_n, \breve{\V{a}}^{(j)}_{m,n} \big) }{\mu^{(j)}_\text{FA} \ist f_{\text{FA}}\big( \V{z}^{(j)}_{m,n} \big)} 
\ist, &\rmv\rmv\rmv\rmv\rmv\rmv b^{(j)}_{m,n} \rmv=\rmv 0 \nn
\end{cases}
\end{align}
and $h\big( \V{x}_n, \breve{\V{a}}^{(j)}_{m,n} , 0, b^{(j)}_{m,n}; \V{z}^{(j)}_{n} \big) \rmv\triangleq\rmv f_{\text{D}}\big(\breve{\V{a}}^{(j)}_{m,n}\big)$. 
The mean number $\mu_{\text{n}, n}^{(j)}$ and the conditional pdf $f_{\text{n},n}\big(\breve{\V{a}}^{(j)}_{m,n}\big|\V{x}_n\big)$ of newly detected features can either be pre-specified 
(e.g., constant with respect to time $n$ and with a spatial distribution that is uniform in $\breve{\V{a}}^{(j)}_{m,n}$) or inferred online by means of a probability hypothesis density (PHD) filter, as described in \cite{Horridge2011PHD}.

\subsection{Joint Posterior pdf}
\label{sec:derivationFactorGraph}

\begin{figure*}[t!]
\centering
\includegraphics[scale=0.92]{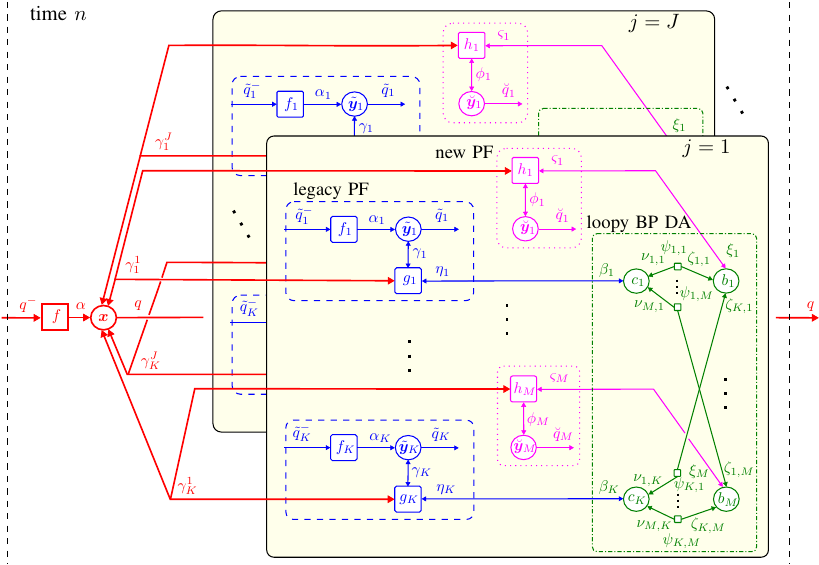}
\vspace{4mm}
\caption{Factor graph representing
the factorization of the joint posterior pdf in \eqref{eq:factorization_post}. The subgraphs corresponding to individual PAs are indicated by boxes with light yellow background. 
All the factor nodes, variable nodes, and messages related to the agent state are represented in red color, those related to the legacy PF states are represented by the blue parts contained in dashed boxes, those related to the new PF states are represented by the magenta parts contained in dotted boxes, and those related to loopy BP DA are represented by the green parts contained in dashed-dotted boxes. The following short notations are used:
$K \rmv\triangleq\rmv K_n^{(j)}$,
$M \rmv\triangleq\rmv M_n^{(j)}$,
$\V{x} \rmv\triangleq \V{x}_n$, 
$\tilde{\V{y}}_k \rmv\triangleq\rmv \tilde{\V{y}}_{k,n}^{(j)}$,
$\breve{\V{y}}_m \rmv\triangleq\rmv \breve{\V{y}}_{m,n}^{(j)}$,
$b_m \rmv\triangleq\rmv b_{m,n}^{(j)}$,
$c_k\rmv\triangleq\rmv c_{k,n}^{(j)}$, 
$\psi_{k,m} \rmv\triangleq\rmv \psi\big(c_{k,n}^{(j)},b_{m,n}^{(j)}\big)$,
$g_k \rmv\triangleq\rmv g\big( \V{x}_n, \tilde{\V{a}}^{(j)}_{k,n} , \tilde{r}^{(j)}_{k,n}, c^{(j)}_{k,n}; \V{z}^{(j)}_{n} \big)$,
$h_m \rmv\triangleq\rmv h\big( \V{x}_n, \breve{\V{a}}_{m,n}^{(j)}, \breve{r}_{m,n}^{(j)}, b_{m,n}^{(j)}; \V{z}_{n}^{(j)} \big)$, 
$f \rmv\triangleq\rmv f(\V{x}_n | \V{x}_{n-1})$,
$f_k \rmv\triangleq\rmv f\big(\tilde{\V{y}}_{k,n}^{(j)} \big| \V{y}_{k,n-1}^{(j)}\big)$,
$\alpha \rmv\triangleq\rmv \alpha( \V{x}_n)$,
$\alpha_k \rmv\triangleq\rmv \alpha_k\big( \tilde{\V{a}}^{(j)}_{k,n} , \tilde{r}^{(j)}_{k,n} \big)$,
$q^- \rmv\triangleq\rmv q(\V{x}_{n-1})$,
$q \rmv\triangleq\rmv q(\V{x}_{n})$,
$\tilde{q}^-_k \rmv\triangleq\rmv \tilde{q}\big( \V{a}^{(j)}_{k,n-1} , r^{(j)}_{k,n-1}\big)$, 
$\tilde{q}_k \rmv\triangleq\rmv \tilde{q}\big( \tilde{\V{a}}^{(j)}_{k,n} , \tilde{r}^{(j)}_{k,n} \big)$, 
$\breve{q}_m \rmv\triangleq\rmv \breve{q}\big(\breve{\V{a}}_{m,n}^{(j)}, \breve{r}_{m,n}^{(j)} \big)$,   
$\beta_{k} \rmv\triangleq\rmv \beta\big(c_{k,n}^{(j)}\big)$,
$\xi_m \rmv\triangleq\rmv \xi\big(b_{m,n}^{(j)}\big)$,
$\eta_{k} \rmv\triangleq\rmv \eta\big(c_{k,n}^{(j)}\big)$, 
$\varsigma_m \rmv\triangleq\rmv \varsigma\big(b_{m,n}^{(j)}\big)$, 
$\nu_{m,k} \rmv\triangleq\rmv \nu^{(p)}_{m \rightarrow k}\big(c_{k,n}^{(j)}\big)$,
$\zeta_{k,m} \rmv\triangleq\rmv \zeta^{(p)}_{k \rightarrow m}\big(b_{m,n}^{(j)}\big)$,
$\gamma_k^j \rmv\triangleq\rmv \gamma_k^{(j)}( \V{x}_n)$, 
$\gamma_k \rmv\triangleq \gamma\big(\tilde{\V{a}}_{k,n}^{(j)},\tilde{r}_{k,n}^{(j)}\big)$,
and
$\phi_m \rmv\triangleq\rmv  \phi\big(\breve{\V{a}}_{m,n}^{(j)},\breve{r}_{m,n}^{(j)}\big)$.}
\label{fig:jpdaFGmod}
\vspace*{-1mm}
\end{figure*}

Using Bayes' rule and independence assumptions related to the state-transition pdfs (see Section~\ref{sec:vec_description}), the prior pdfs (see Section~\ref{eq:JointAssociation}), 
and the likelihood model (see Section~\ref{sec:meas_likelihood}), the joint posterior pdf of $\V{x}_{n'}$, $\tilde{\V{y}}_{n'}$, $\breve{\V{y}}_{n'}$, $\V{c}_{n'}$, $\V{b}_{n'}$, and $\V{m}_{n'}$ for all $n' \rmv= 1,\ldots,n$ is obtained as (by using \eqref{eq:state_space}, \eqref{eq:allTimeslhf}, and \eqref{eq:allTimesPrior})
\begin{align}
&f( \V{x}_{1:n}, \V{y}_{1:n}, \V{c}_{1:n}, \V{b}_{1:n} ,\V{m}_{1:n} | \V{z}_{1:n} ) \nn \\[1mm]
&\propto f( \V{z}_{1:n} | \V{x}_{1:n}, \V{y}_{1:n},  \V{c}_{1:n}, \V{m}_{1:n}) f( \V{x}_{1:n}, \V{y}_{1:n}, \V{c}_{1:n} ,\V{m}_{1:n} ) \nn \\[-1mm]
&=  f(\V{x}_{1})  \Bigg( \prod^{J}_{j'=1} \rmv f\big( \V{z}^{(j')}_{1} \big| \V{x}_{1}, \breve{\V{y}}^{(j')}_{1} \rmv, \V{c}^{(j')}_{1}\rmv, M_1^{(j')}\big) \nn \\[-1.5mm]
&\hspace{2mm} \times  f\big(\breve{\V{y}}^{(j')}_{1} \rmv, \V{c}^{(j')}_{1} \rmv,  M^{(j')}_{1} \big|\V{x}_1 \big)  \rmv \Bigg) 
   \rmv\prod^{n}_{n'=2} \! f(\V{x}_{n'}|\V{x}_{n'-1}) \nn \\[-1.2mm]
&\hspace{2mm} \times \prod^{J}_{j=1} \Bigg(\prod_{k=1}^{K_{n-1}^{(j)}}\rmv\rmv\rmv f\big(\tilde{\V{y}}^{(j)}_{k,n'} \big| \V{y}^{(j)}_{k,n'-1}\big)\Bigg) \nn \\[-.5mm]
&\hspace{2mm}\times  f\big( \V{z}^{(j)}_{n'} \big| \V{x}_{n'}, \tilde{\V{y}}^{(j)}_{n'} \rmv, \breve{\V{y}}^{(j)}_{n'} \rmv, \V{c}^{(j)}_{n'} \rmv, M^{(j)}_{n'}\big) \ist f\big(\breve{\V{a}}^{(j)}_{n'} \big|\V{x}_{n'}, \breve{\V{r}}^{(j)}_{n'} \rmv,M^{(j)}_{n'}\big) \nn \\[1mm]
&\hspace{2mm}\times  p\big(\V{c}^{(j)}_{n'} \rmv, \V{b}^{(j)}_{n'} \rmv, \breve{\V{r}}^{(j)}_{n'}\rmv, M^{(j)}_{n'} \big|\V{x}_{n'}, \tilde{\V{y}}^{(j)}_{n'}\big) \ist. 
\label{eq:factorization_post1}
\end{align}
After inserting expression \eqref{eq:factorGraph} and performing some simple manipulations, Eq. \eqref{eq:factorization_post1} can be reformulated as
\begin{align}
&f( \V{x}_{1:n}, \V{y}_{1:n}, \V{c}_{1:n}, \V{b}_{1:n} ,\V{m}_{1:n} | \V{z}_{1:n} ) \nn \\[-1.3mm]
&\propto  f(\V{x}_{1}) \Bigg(\prod^{J}_{j'=1} \rmv \prod^{M^{(j')}_{1}}_{m'=1} \! h\big( \V{x}_{1}, \breve{\V{a}}^{(j')}_{m'\!,1} , \breve{r}^{(j')}_{m'\!,1}, b^{(j')}_{m'\!,1}; \V{z}^{(j')}_{1} \big) \! \Bigg) \nn \\
&\hspace{4mm}\times \prod^{n}_{n'=2}  \! f(\V{x}_{n'}|\V{x}_{n'-1}) \prod^{J}_{j=1} \rmv\Psi\big(\V{c}^{(j)}_{n'} \rmv,\V{b}^{(j)}_{n'}\big)\nn\\[0mm]
&\hspace{4mm}\times \rmv \Bigg( \prod^{K^{(j)}_{n'-1}}_{k=1}\rmv\rmv\rmv f\big(\tilde{\V{y}}^{(j)}_{k,n'} \big| \V{y}^{(j)}_{k,n'-1}\big) g\big( \V{x}_{n'}, \tilde{\V{a}}^{(j)}_{k,n'} , \tilde{r}^{(j)}_{k,n'}, c^{(j)}_{k,n'}; \V{z}^{(j)}_{n'} \big)\rmv\Bigg) \nn \\
&\hspace{4mm}\times \prod^{M^{(j)}_{n'}}_{m=1} h\big( \V{x}_{n'}, \breve{\V{a}}^{(j)}_{m,n'} , \breve{r}^{(j)}_{m,n'}, b^{(j)}_{m,n'}; \V{z}^{(j)}_{n'} \big)  \ist.
\label{eq:factorization_post}\\[-6mm]\nn
\end{align}
This factorization is represented by the factor graph \cite{Loeliger2004SPM,KschischangTIT2001} shown in Fig.~\ref{fig:jpdaFGmod}. 

\section{The BP-SLAM Algorithm}\label{sec:BPmessgepassing}
The proposed BP-SLAM algorithm performs Bayesian detection and estimation of the relevant states. The required posterior distributions are calculated in an efficient, time-recursive manner by using BP message passing \cite{KschischangTIT2001} on the factor graph in Fig.~\ref{fig:jpdaFGmod}.

\subsection{Detection and Estimation}\label{sec:problem}

Our goal is to estimate the agent state $\V{x}_{n}$ and to detect and estimate the PF states $\V{a}_{k,n}^{(j)}$ from the total measurement vector $\V{z}_{1:n}$. For estimating $\V{x}_n$, we will develop an approximate calculation of the minimum mean-square error (MMSE) estimator \cite{kay1993}
\begin{equation}
\hat{\V{x}}^\text{MMSE}_{n} \,\triangleq\rmv \int\rmv \V{x}_n \ist f(\V{x}_n |\V{z}_{1:n}) \ist \mathrm{d}\V{x}_n \,.
\label{eq:mmse_agent}
\end{equation}
This estimator involves the posterior pdf $f(\V{x}_n | \V{z}_{1:n})$. Furthermore, detecting (i.e., determining the existence of) PF $k \in \Set{K}^{(j)}_n$ at time $n$ is based on the posterior existence probability $p\big({r}_{k,n}^{(j)} \!=\!1 \big| \V{z}_{1:n}\big)$. This probability can be obtained from the posterior pdf of the augmented PF state, $f\big(\V{y}_{k,n}^{(j)}\big|\V{z}_{1:n}\big) \rmv=\rmv  f\big(\V{a}_{k,n}^{(j)}\ist, r_{k,n}^{(j)}\big|\V{z}_{1:n}\big)$, by a marginalization, i.e.,
\begin{align}
p\big({r}_{k,n}^{(j)} \!=\!1 \big| \V{z}_{1:n}\big) =\rmv \int \rmv f\big({\V{a}}_{k,n}^{(j)}\ist, r_{k,n}^{(j)} \!=\! 1\big|\V{z}_{1:n}\big) \ist\mathrm{d}{\V{a}}_{k,n}^{(j)} \ist.
\label{eq:exist_prob_PFs}
\\[-5mm]\nn
\end{align}
Then PF $k$ is defined to be detected at time $n$ if $p\big({r}_{k,n}^{(j)} \!=\!1 \big| \V{z}_{1:n}\big) \rmv>\rmv P_\text{det}$, where $P_\text{det}$ is a detection threshold. 
The states $\V{a}_{k,n}^{(j)}$ of the detected PFs are finally estimated as
\begin{align}
\hat{\V{a}}^{(j)\ist\text{MMSE}}_{k,n} \,\triangleq\rmv \int \rmv {\V{a}}_{k,n}^{(j)}\, f\big({\V{a}}_{k,n}^{(j)}\ist \big|{r}_{k,n}^{(j)} \!=\! 1,\V{z}_{1:n}\big) \ist \mathrm{d}{\V{a}}_{k,n}^{(j)} \ist,
\label{eq:MMSE_PFs}
\\[-6mm]\nn
\end{align}
where
\begin{align}
\nn\\[-8mm]
f\big({\V{a}}_{k,n}^{(j)}\ist \big|{r}_{k,n}^{(j)} \!=\! 1,\V{z}_{1:n}\big)  = \frac{f\big({\V{a}}_{k,n}^{(j)}\ist,{r}_{k,n}^{(j)} \!=\! 1\big|\V{z}_{1:n}\big)}{p\big({r}_{k,n}^{(j)} \!=\!1 \big| \V{z}_{1:n}\big)}\ist.\label{eq:marginal_PFs}
\\[-5mm]\nn
\end{align}

The posterior existence probabilities $p\big({r}_{k,n}^{(j)} \!=\!1 \big| \V{z}_{1:n}\big)$ in \eqref{eq:exist_prob_PFs} are also used in a different context.
To prevent an indefinite increase of the total number of PFs for PA $j$ due to \eqref{eq:numandsetFeat}, i.e., $K_{n}^{(j)}\rmv=\rmv K_{n-1}^{(j)} + M_{n}^{(j)}$, a pruning of the PFs is employed. More specifically, PF $k$ is retained only if $p\big(r_{k,n}^{(j)} \!=\! 1\big| \V{z}_{1:n}\big)$ exceeds a suitably chosen pruning threshold $P_\text{prun}\rmv$. 

\subsection{Message Passing Algorithm}\label{sec:BP_algorithm}

The posterior pdfs $f(\V{x}_n |\V{z}_{1:n})$ and $f\big({\V{a}}_{k,n}^{(j)}\ist ,{r}_{k,n}^{(j)}\big|\V{z}_{1:n}\big)$ involved in \eqref{eq:mmse_agent}--\eqref{eq:marginal_PFs} are marginal pdfs of the joint posterior pdf $f( \V{x}_{1:n}, \V{y}_{1:n}, \V{c}_{1:n}, \V{b}_{1:n},\V{m}_{1:n} | \V{z}_{1:n} )$ in \eqref{eq:factorization_post}. Because a direct marginalization is infeasible, we use loopy (iterative) BP message passing \cite{KschischangTIT2001} on the factor graph in Fig.~\ref{fig:jpdaFGmod}. Since the factor graph contains loops, the resulting beliefs are only approximations of the respective posterior pdfs, and there is no canonical order in which the messages should be computed \cite{KschischangTIT2001}.
In our method, we choose the order according to the following rules: 
(i) Messages are not passed backward in time; 
(ii) \emph{iterative} message passing is only performed for DA, and only for each time step and for each PA separately (i.e., 
in particular, for the loops connecting different PAs, we only perform a single message passing iteration); 
(iii) along an edge connecting an agent state variable node and a new PF state variable node, messages are only sent from the former to the latter. 
The resulting BP message passing algorithm is presented in what follows;
cf.\ also the underlying factor graph in Fig.~\ref{fig:jpdaFGmod}. We note that similarly to the ``dummy pdfs'' introduced in Section~\ref{sec:vec_description}, we consider BP messages 
$\varphi\big(\V{y}^{(j)}_{k,n}\big) \rmv=\rmv \varphi\big(\V{a}^{(j)}_{k,n}\ist, r^{(j)}_{k,n}\big)$ also for the non-existing PF states, i.e., for $r^{(j)}_{k,n} \!=\rmv 0$.
We define these messages by setting $\varphi\big(\V{a}^{(j)}_{k,n}\ist, 0\big) = \varphi^{(j)}_{k,n}$ (note that these messages are not pdfs and thus are not required to integrate to $1$).

First, a prediction step is performed. The prediction message for the agent state is given by
\be
 \alpha( \V{x}_n) =\rmv \int \rmv f(\V{x}_n|\V{x}_{n-1}) \ist q(\V{x}_{n-1}) \ist \mathrm{d}\V{x}_{n-1} \ist,
  \label{eq:stateTransitionMessageAgent}
\ee
and the prediction message for the legacy PFs is given by
\begin{align}
\alpha_k\big( \tilde{\V{a}}_{k,n}^{(j)} \ist, \tilde{r}_{k,n}^{(j)}\big) &=\rmv  \sum_{r^{(j)}_{k,n-1} \in \{0,1\}}\int \rmv f\big(\tilde{\V{a}}_{k,n}^{(j)}\ist, \tilde{r}_{k,n}^{(j)} \big| \V{a}_{k,n-1}^{(j)}\ist, r^{(j)}_{k,n-1} \big) \nn \\
&\hspace{5mm}\times\tilde{q}\big(\V{a}_{k,n-1}^{(j)}\ist, r^{(j)}_{k,n-1} \big) \ist \mathrm{d}\V{a}_{k,n-1}^{(j)} \ist,
 \label{eq:stateTransitionMessageFeature} 
\end{align}
$k \rmv\rmv\in\rmv\rmv \mathcal{K}_{k-1}^{(j)}$, where the beliefs of the mobile agent state, $q(\V{x}_{n-1})$,\linebreak and of the PF states, $\tilde{q}\big(\V{a}_{k,n-1}^{(j)}\ist, r^{(j)}_{k,n-1} \big)$, were calculated at the preceding time $n \rmv-\! 1$. Inserting \eqref{eq:survival_transition} and \eqref{eq:dummy_transition} for $f\big(\tilde{\V{a}}_{k,n}^{(j)}\ist, \tilde{r}_{k,n}^{(j)} \big| \V{a}_{k,n-1}^{(j)}\ist, r^{(j)}_{k,n-1}\rmv\rmv=\rmv\rmv 1\big)$ and $f\big(\tilde{\V{a}}_{k,n}^{(j)}\ist, \tilde{r}_{k,n}^{(j)} \big| \V{a}_{k,n-1}^{(j)}\ist,$\linebreak$ r^{(j)}_{k,n-1}\rmv=\rmv 0\big)$, respectively, we obtain for $\tilde{r}_{k,n}^{(j)} \!=\! 1$
\be
 \alpha_k\big( \tilde{\V{a}}_{k,n}^{(j)}\ist, 1\big) 
   =\ist P_\text{s} \rmv \int \rmv f\big(\tilde{\V{a}}_{k,n}^{(j)} \big| \V{a}_{k,n-1}^{(j)}\big) \ist \tilde{q}\big(\V{a}_{k,n-1}^{(j)}\ist, 1 \big) \ist \mathrm{d}\V{a}_{k,n-1}^{(j)} \ist, 
 \label{eq:stateTransitionMessageFeature_r1}
\vspace{-1mm}
\ee
and for $\tilde{r}_{k,n}^{(j)} \!=\rmv 0$
\be
 \alpha_{k,n}^{(j)} = (1 \!-\rmv P_\text{s}\big) \rmv\int\rmv \tilde{q}\big(\V{a}_{k,n-1}^{(j)}\ist, 1 \big) \ist \mathrm{d} \V{a}_{k,n-1}^{(j)} +\ist 	\tilde{q}^{(j)}_{k,n-1} \ist,
 \label{eq:stateTransitionMessageFeature_r0}
\vspace{1mm}
\ee
where $\alpha_{k,n}^{(j)} \triangleq \int \rmv\alpha_k\big( \tilde{\V{a}}_{k,n}^{(j)}\ist, 0\big) \ist \mathrm{d}\tilde{\V{a}}_{k,n}^{(j)}$ and $\tilde{q}^{(j)}_{k,n-1} \triangleq \int \rmv \tilde{q}\big( \V{a}_{k,n-1}^{(j)}\ist, 0\big) \ist \mathrm{d}\V{a}_{k,n-1}^{(j)}\ist$.

After the prediction step, the following calculations are performed for all legacy PFs $k \!\in\! \Set{K}_{n-1}^{(j)}$ and for all new PFs $m \!\in\! \Set{M}_{n}^{(j)}$, for all PAs $j \rmv\in\rmv \{1,\ldots,J\}$ in parallel:
\begin{enumerate}

\item \textit{Measurement evaluation for legacy PFs}: The messages $\beta\big( c_{k,n}^{(j)} \big)$ passed to the variable nodes corresponding to the feature-oriented 
DA variables $c^{(j)}_{k,n}$ (cf. Fig.~\ref{fig:jpdaFGmod}) are calculated as
\begin{align}
\nn\\[-5mm]
\beta\big( c_{k,n}^{(j)} \big) &\rmv\rmv\rmv=\rmv\rmv\rmv \int\!\!\!\int \rmv\rmv\rmv \alpha_k\big( \tilde{\V{a}}_{k,n}^{(j)},1\big) \ist \alpha(\V{x}_n) \ist g\big( \V{x}_n, \tilde{\V{a}}_{k,n}^{(j)}, 1, c_{k,n}^{(j)};\V{z}_{n}^{(j)} \big) \nn \\
&\rmv\rmv\rmv \hspace{5mm}\times \mathrm{d}\V{x}_{n} \ist \mathrm{d}\tilde{\V{a}}_{k,n}^{(j)} \ist+\ist 1\big(c_{k,n}^{(j)}\big)\ist\alpha_{k,n}^{(j)} \ist. 
\label{eq:bp_measevalutionLF}
\\[-5mm]\nn
\end{align}

\item \textit{Measurement evaluation for new PFs}: The messages $\xi\big(b^{(j)}_{m,n}\big)$ passed to the variable nodes corresponding to the measurement-oriented DA variables $b^{(j)}_{m,n}$ are calculated as
\begin{align}
\xi\big(b^{(j)}_{m,n}\big) &=\rmv\rmv \sum_{\breve{r}^{(j)}_{m,n} \in \{0,1\}} \int\!\!\!\int \rmv h\big( \V{x}_{n}, \breve{\V{a}}^{(j)}_{m,n}, \breve{r}^{(j)}_{m,n}, b_{m,n}^{(j)} ; \V{z}^{(j)}_{n} \big) \nn \\
&\hspace{5mm}\times  \alpha(\V{x}_n) \, \mathrm{d}\V{x}_{n} \ist \mathrm{d}\breve{\V{a}}^{(j)}_{m,n} \ist.
\label{eq:bp_measevalutionNF1}\\[-6.5mm]
\nn
\end{align}
Using the expression of $h\big( \V{x}_{n}, \breve{\V{a}}^{(j)}_{m,n}, \breve{r}^{(j)}_{m,n}, b^{(j)}_{m,n}; \V{z}^{(j)}_{n} \big)$ stated in Section \ref{sec:modPrior}, 
Eq.\ \eqref{eq:bp_measevalutionNF1} is easily seen to simplify to $\xi\big(b^{(j)}_{m,n}\big) \!=\! 1$
for $b^{(j)}_{m,n} \!\in\! \Set{K}^{(j)}_{n-1}$, and for $b^{(j)}_{m,n} \!=\rmv 0$ it becomes
\begin{align}
\xi\big(b^{(j)}_{m,n}\big) &= 1 + \frac{ \mu_{\text{n}, n}^{(j)} }{\mu_\text{FA}^{(j)} \ist f_{\text{FA}}\big( \V{z}^{(j)}_{m,n} \big)} 
  \int\!\!\!\int\rmv \alpha(\V{x}_n) \ist f_{\text{n},n}\big(\breve{\V{a}}^{(j)}_{m,n}\big| \V{x}_n \big) \nn \\[1.5mm]
&\hspace{5mm}\times  f\big( \V{z}^{(j)}_{m,n} \big|\V{x}_{n}, \breve{\V{a}}^{(j)}_{m,n} \big) \, \mathrm{d}\V{x}_{n} \ist \mathrm{d}\breve{\V{a}}^{(j)}_{m,n} \ist.
\label{eq:bp_measevalutionNF}
\end{align}

\item \textit{Iterative data association}: Next, from $\beta\big( c_{k,n}^{(j)} \big)$ and $\xi\big(b^{(j)}_{m,n}\big)$, messages 
$\eta\big( c_{k,n}^{(j)} \big)$ and $\varsigma\big( b_{m,n}^{(j)} \big)$ are obtained by means of loopy (iterative) BP.
First, for each measurement $m \!\in\! \Set{M}_n^{(j)}$, messages $\nu_{m\rightarrow\mpcidxsym}^{(p)}\big(c_{\mpcidxsym,\timestepsym}\anchoridx\big)$ 
and $\zeta_{\mpcidxsym \rightarrow m}^{(p)}\big(b_{m,\timestepsym}\anchoridx\big)$ are calculated iteratively according to \cite{WilliamsTAE2014,MeyerTSP2017}
\begin{align}
\nu_{m\rightarrow\mpcidxsym}^{(p)}\big(c_{\mpcidxsym,\timestepsym}\anchoridx\big) 
&=\! \sum^{K_{n-1}^{(j)}}_{b_{m,\timestepsym}\anchoridx = 0} \!\!\xi\big( b_{m,n}^{(j)} \big) \ist \psi\big(c_{\mpcidxsym,\timestepsym}\anchoridx, b_{m,\timestepsym}\anchoridx\big)  \nn \\[-1mm]
&\hspace{5mm}\times\rmv\rmv\rmv\rmv\prod_{\mpcidxsym' \in \Set{K}^{(j)}_{n-1}\backslash\{k\}} \!\!\! \zeta_{k' \rightarrow m}^{(p-1)}\big(b_{m,n}\anchoridx\big)
\label{eq:featureDARV}\\
\zeta_{\mpcidxsym \rightarrow m}^{(p)}\big(b_{m,\timestepsym}\anchoridx\big) 
&=\! \sum^{M_n^{(j)}}_{c_{\mpcidxsym,\timestepsym}\anchoridx = 0} \!\!\beta\big( c_{\mpcidxsym,\timestepsym}\anchoridx \big) \ist 
  \psi\big(c_{\mpcidxsym,\timestepsym}\anchoridx, b_{m,\timestepsym}\anchoridx\big)  \nn \\[-1mm] 
&\hspace{5mm}\times\rmv\rmv\rmv\rmv\prod_{m' \in \Set{M}_n^{(j)}\backslash\{m\}} \!\!\! \nu_{m'\rightarrow\mpcidxsym}^{(p)}\big(c_{\mpcidxsym,\timestepsym}\anchoridx\big) \ist, 
\label{eq:measurementDARV} \\[-5mm]
\nn
\end{align}
for $k \rmv\in\rmv \mathcal{K}_{n-1}^{(j)}$, $m \rmv\in\rmv \mathcal{M}_{n}^{(j)}\ist$, and iteration index $p \rmv=\rmv 1,\ldots,$\linebreak$ P$. The recursion defined by \eqref{eq:featureDARV} and \eqref{eq:measurementDARV} is initialized (for $p \!=\!  0$) by
$\zeta_{\mpcidxsym \rightarrow m}^{(0)}\big(b_{m,\timestepsym}\anchoridx\big) \rmv =\sum^{M_n^{(j)}}_{c_{\mpcidxsym,\timestepsym}\anchoridx = 0}  \beta\big( c_{\mpcidxsym,\timestepsym}\anchoridx \big) $\linebreak$\times\psi\big(c_{\mpcidxsym,\timestepsym}\anchoridx, b_{m,\timestepsym}\anchoridx\big)$. Then, after the last iteration $p \rmv=\rmv P\rmv$, the messages $\eta\big( c_{k,n}^{(j)} \big)$ and $\varsigma\big( b_{m,n}^{(j)} \big)$ are calculated as
\begin{align}
	\eta\big( c_{\mpcidxsym,\timestepsym}\anchoridx \big) &=\! \prod_{m \in \Set{M}_n^{(j)}} \!\!\! \nu_{m\rightarrow\mpcidxsym}^{(P)}\big(c_{\mpcidxsym,\timestepsym} \anchoridx\big) \\[1mm] 
	\varsigma\big( b_{m,\timestepsym}\anchoridx \big) &=\! \prod_{k \in \Set{K}_{n-1}^{(j)}} \!\!\rmv \zeta_{\mpcidxsym \rightarrow m}^{(P)}\big(b_{m,\timestepsym}\anchoridx\big) \ist.
	\label{eq:messagesDA}\\[-7mm]\nn
\end{align}

\item \textit{Measurement update for the agent}: From $\eta\big( c_{k,n}^{(j)} \big)$, $\alpha_k\big( \tilde{\V{a}}_{k,n}^{(j)}, 1\big)$, and $ \alpha_{k,n}^{(j)}\ist$, 
the message $\gamma^{(j)}_k(\V{x}_n)$ related to the agent is calculated as
\begin{align}
\hspace{-2mm}\gamma^{(j)}_k(\V{x}_n) &=\! \sum^{M_n^{(j)}}_{c_{\mpcidxsym,\timestepsym}\anchoridx = 0} \!\! \eta\big( c_{k,n}^{(j)} \big) 
\rmv\int \! g\big( \V{x}_n, \tilde{\V{a}}_{k,n}^{(j)}, 1,c_{k,n}^{(j)}; \V{z}_{n}^{(j)} \big)\nn \\
&\hspace{2mm}\times \alpha_k\big( \tilde{\V{a}}_{k,n}^{(j)} , 1\big) \mathrm{d} \tilde{\V{a}}_{k,n}^{(j)} 
\ist+\ist \eta\big( c_{k,n}^{(j)} \!\rmv=\! 0\big) \ist\alpha_{k,n}^{(j)} \ist. 
\label{eq:measurementUpdateAgent}\\[-4mm]\nn
\end{align}
		
\item \textit{Measurement update for legacy PFs}: Similarly, the messages $\gamma\big(\tilde{\V{a}}_{k,n}^{(j)},\tilde{r}_{k,n}^{(j)}\big)$ related to the legacy PFs
are calculated as
\begin{align}
\label{eq:measurementUpdateLegacy1}
\gamma\big(\tilde{\V{a}}_{k,n}^{(j)},1\big) &=\! \sum^{M_n^{(j)}}_{c_{\mpcidxsym,\timestepsym}\anchoridx = 0} \!\! \eta\big( c_{k,n}^{(j)} \big) \rmv\int \! g\big( \V{x}_n, \tilde{\V{a}}_{k,n}^{(j)}, 1,c_{k,n}^{(j)}; \V{z}_{n}^{(j)} \big)  \nn \\ 
&\hspace{5mm}\times \ist \alpha(\V{x}_n) \mathrm{d}\V{x}_n \\[1.5mm]
\gamma_{k,n}^{(j)} &\triangleq\ist \gamma\big(\tilde{\V{a}}_{k,n}^{(j)},0\big) = \eta\big( c_{k,n}^{(j)} \!\rmv=\! 0\big) \ist.
\label{eq:measurementUpdateLegacy2}
\end{align}

\item \textit{Measurement update for new PFs}: Finally, the messages $\phi\big(\breve{\V{a}}_{m,n}^{(j)},\breve{r}_{m,n}^{(j)}\big)$ related to the new PFs
are calculated as
\begin{align}
\label{eq:measurementUpdateNew1}
\phi\big(\breve{\V{a}}_{m,n}^{(j)},1\big) &=\ist \varsigma\big( b_{m,n}^{(j)} \!\rmv=\! 0\big) \rmv\int\!  h\big( \V{x}_n, \breve{\V{a}}_{m,n}^{(j)}, 1,0; \V{z}_{n}^{(j)} \big)  \, \nn \\
&\hspace{5mm}\times
  \ist \alpha(\V{x}_n) \mathrm{d}\V{x}_n \\[.5mm] 
\phi_{m,n}^{(j)} &\triangleq\ist \phi\big(\breve{\V{a}}_{k,n}^{(j)},0\big) =\rmv \sum^{K_{n-1}^{(j)}}_{b_{m,\timestepsym}\anchoridx = 0} \!\! \varsigma\big( b_{m,n}^{(j)}\big) \ist.
\label{eq:measurementUpdateNew2}
\end{align}
\end{enumerate}

Once these messages are available, the beliefs approximating the desired marginal posterior pdfs are obtained. The belief for the agent state is given, up to a normalization factor, 
by 
\be
q(\V{x}_n) \ist\propto\ist \alpha( \V{x}_n) \prod_{j=1}^J \prod_{k \in \Set{K}_{n-1}^{(j)}} \!\!\rmv \gamma^{(j)}_k( \V{x}_n) \ist.
\label{eq:q_x} 
\ee
The belief $q(\V{x}_n)$ provides an approximation of the marginal posterior pdf $f(\V{x}_n |\V{z}_{1:n})$, and it is used instead of $f(\V{x}_n |\V{z}_{1:n})$ in \eqref{eq:mmse_agent}.
Furthermore, the beliefs $\tilde{q}\big(\tilde{\V{a}}_{k,n}^{(j)}, \tilde{r}_{k,n}^{(j)}\big)$ for the augmented states of the legacy PFs, 
$\tilde{\V{y}}_{k,n}^{(j)} \rmv=\rmv \big[\tilde{\V{a}}_{k,n}^{(j)\ist\text{T}} \; \tilde{r}_{k,n}^{(j)}\big]\trans\rmv$, are calculated as 
\begin{align}
\label{eq:beliefLegacy1}
\tilde{q}\big(\tilde{\V{a}}_{k,n}^{(j)},1\big) &\ist\propto\ist \alpha_k\big( \tilde{\V{a}}_{k,n}^{(j)}, 1\big) \ist \gamma\big( \tilde{\V{a}}_{k,n}^{(j)},1\big)\\[1mm]
\tilde{q}_{k,n}^{(j)} &\triangleq\, \tilde{q}\big(\tilde{\V{a}}_{k,n}^{(j)},0\big) \ist\propto\ist \alpha_{k,n}^{(j)} \ist \gamma_{k,n}^{(j)}\ist,
\label{eq:beliefLegacy2}
\end{align}
and the beliefs $\breve{q}\big(\breve{\V{a}}_{m,n}^{(j)}, \breve{r}_{m,n}^{(j)}\big)$ for the augmented states of the new PFs, 
$\breve{\V{y}}_{m,n}^{(j)} =  \big[\breve{\V{a}}_{m,n}^{(j)\ist\text{T}} \; \breve{r}_{m,n}^{(j)}\big]\trans\rmv$, as 
\begin{align}
\label{eq:beliefNew1}
\breve{q}\big(\breve{\V{a}}_{m,n}^{(j)},1\big) &\ist\propto\ist \phi\big(\breve{\V{a}}_{m,n}^{(j)},1\big) \\
\breve{q}_{m,n}^{(j)} &\triangleq\, \breve{q}\big(\breve{\V{a}}_{m,n}^{(j)},0\big) \ist\propto\ist \phi_{m,n}^{(j)} \ist.
\label{eq:beliefNew2}
\end{align}
In particular, $\tilde{q}\big(\tilde{\V{a}}_{k,n}^{(j)},1\big)$ and $\breve{q}\big(\breve{\V{a}}_{m,n}^{(j)},1\big)$ approximate the marginal posterior pdf 
$f\big(\V{a}_{k'\!,n}^{(j)}, r_{k'\!,n}^{(j)} \!=\! 1\big|\V{z}_{1:n}\big)$, where $k' \!\in \Set{K}_{n-1}^{(j)} \cup \Set{M}_{n}^{(j)}$ (assuming an appropriate index mapping 
between $k$, $m$ on the one hand and $k'$ on the other), and they are used in \eqref{eq:exist_prob_PFs}--\eqref{eq:marginal_PFs}.

The BP-SLAM algorithm is summarized in Algorithm 1. A flowchart is available online at 
https://www2.spsc.tugraz.\linebreak at/people/eriklei/BP-SLAM/
and also at 
https://gitlab.com/\linebreak erikleitinger/BP-Multipath-basedSLAM.
A computationally feasible sequential Monte Carlo (particle-based) implementation can be obtained via the approach in \cite{MeyerTSPIN2016,MeyerTSP2017}. In our case, the sequential Monte Carlo implementation uses a ``stacked state'' \cite{MeyerTSPIN2016} comprising the agent state and the PF states. The resulting complexity scales only linearly in the number of particles. 
MATLAB code for this particle-based implementation is available online at the addresses indicated above.

\vspace*{5.7mm}

{
\hrule
\vspace*{-1.8mm}
\begin{center}
{\small\sc Algorithm 1:\, BP-SLAM (Recursion from $n\rmv\rmv-\rmv\rmv1$ to $n$)}
\end{center}
\vspace*{-1.35mm}
\hrule 
\vspace*{2.6mm}
\small
\renewcommand{\baselinestretch}{1.15}\normalsize\small
\noindent
\textbf{\textit{Step 1---Prediction}:}\\[.5mm]
\hspace*{2mm} Calculate $\alpha(\V{x}_{n})$ from $q(\V{x}_{n-1})$ according to \eqref{eq:stateTransitionMessageAgent}, 
and calculate $\alpha_k\big( \tilde{\V{a}}_{k,n}^{(j)}\ist, 1\big)$ and $\alpha_{k,n}^{(j)}$ from $\tilde{q}\big(\V{a}_{k,n-1}^{(j)},r_{k,n-1}^{(j)}\big)$, $k \rmv\in\rmv \Set{K}^{(j)}_{n-1}\ist$, $j \rmv\in\rmv \{1,$\linebreak$\ldots,J\}$ according to 
\eqref{eq:stateTransitionMessageFeature_r1}, \eqref{eq:stateTransitionMessageFeature_r0}. \\[1.7mm] 
\textbf{\textit{Step 2---Measurement evaluation}:}\\[.5mm]
\hspace*{2mm} \emph{For legacy PFs}:\, Calculate $\beta\big(c^{(j)}_{k,n}\big)$, $k \rmv\in\rmv \Set{K}^{(j)}_{n-1}\ist$, $j \rmv\in\rmv \{1,\ldots,J\}$ according to \eqref{eq:bp_measevalutionLF}.\\[.5mm]
\hspace*{2mm} \emph{For new PFs}:\, Calculate $\xi\big(b^{(j)}_{m,n}\big)$, $m \rmv\in\rmv \Set{M}^{(j)}_n\!$, $j \rmv\in\rmv \{1,\ldots,J\}$ according to \eqref{eq:bp_measevalutionNF1}, 
\eqref{eq:bp_measevalutionNF}.\\[1.7mm] 
\textbf{\textit{Step 3---Iterative data association}:}\\[.5mm] 
\hspace*{2mm} Calculate $\eta\big( c_{k,n}^{(j)} \big)$, $k \rmv\in\rmv \Set{K}^{(j)}_{n-1}\ist$, $j \rmv\in\rmv \{1,\ldots,J\}$  and $\varsigma\big( b_{m,n}^{(j)}\big)$, $m \rmv\in\rmv \Set{M}^{(j)}_n$, $j \rmv\in\rmv \{1,\ldots,J\}$ according to \eqref{eq:featureDARV}--\eqref{eq:messagesDA}. \\[1.7mm] 
\textbf{\textit{Step 4---Measurement update}:}\\[.5mm]
\hspace*{2mm} \emph{For the mobile agent}:\, Calculate $\gamma^{(j)}_k(\V{x}_n)$, $k \rmv\in\rmv \Set{K}^{(j)}_{n-1}\ist$, $j \rmv\in\rmv \{1,\ldots,$\linebreak$J\}$ according to \eqref{eq:measurementUpdateAgent}.\\[.5mm]
\hspace*{2mm} \emph{For legacy PFs}:\, Calculate $\gamma\big(\tilde{\V{a}}_{k,n}^{(j)},1\big)$ and $\gamma_{k,n}^{(j)}$, $k \rmv\in\rmv \Set{K}^{(j)}_{n-1}\ist$, $j \rmv\in\rmv \{1,$\linebreak$\ldots,J\}$ according to \eqref{eq:measurementUpdateLegacy1}, \eqref{eq:measurementUpdateLegacy2}.\\[.5mm]
\hspace*{2mm} \emph{For new PFs}:\, Calculate $\phi\big(\breve{\V{a}}_{m,n}^{(j)},1\big)$ and $\phi_{m,n}^{(j)}$, $m \rmv\in\rmv \Set{M}^{(j)}_n\rmv$, $j \rmv\in\rmv \{1,$\linebreak$\ldots,J\}$ according to \eqref{eq:measurementUpdateNew1}, \eqref{eq:measurementUpdateNew2}.\\[1.7mm] 
\textbf{\textit{Step 5---Belief calculation}:}\\[.5mm]
\hspace*{2mm} \emph{For the mobile agent}:\, Calculate $q(\V{x}_n)$ according to \eqref{eq:q_x}.\\[.5mm]
\hspace*{2mm} \emph{For legacy PFs}:\, Calculate $\tilde{q}\big(\tilde{\V{a}}_{k,n}^{(j)},1\big)$ and $\tilde{q}_{k,n}^{(j)}$, $k \rmv\in\rmv \Set{K}^{(j)}_{n-1}$, $j \rmv\in\rmv \{1,$\linebreak$\ldots,J\}$ 
  according to \eqref{eq:beliefLegacy1}, \eqref{eq:beliefLegacy2}. \\[.5mm]
\hspace*{2mm} \emph{For new PFs}:\, Calculate $\breve{q}\big(\breve{\V{a}}_{m,n}^{(j)},1\big)$ and $\breve{q}_{m,n}^{(j)}$, $m \rmv\in\rmv \Set{M}^{(j)}_n\rmv$, $j \rmv\in\rmv \{1,$\linebreak$\ldots,J\}$ 
  according to \eqref{eq:beliefNew1}, \eqref{eq:beliefNew2}.\\[1.7mm]
\textbf{\textit{Step 6---Pruning}:}\\[1mm]
\hspace*{2mm} Determine the set $\tilde{\Set{K}}^{(j)}_{n} \rmv= \Set{K}^{(j)}_{n-1} \cup \Set{M}^{(j)}_n$.\\[1mm]
\hspace*{2mm} For all $j \rmv\in\rmv \{1,\ldots,J\}$, reinterpret/reindex the beliefs $\tilde{q}\big(\tilde{\V{a}}_{k'\rmv,n}^{(j)},1\big)$ and $\tilde{q}_{k'\rmv,n}^{(j)}$, $k' \!\rmv\in\rmv \Set{K}^{(j)}_{n-1}$ and the beliefs $\breve{q}\big(\breve{\V{a}}_{m,n}^{(j)},1\big)$ and $\breve{q}_{m,n}^{(j)}$, $m \rmv\in\rmv \Set{M}^{(j)}_n$ as beliefs $\tilde{q}\big(\V{a}_{k,n}^{(j)},1\big)$ and $q_{k,n}^{(j)}$ of the PFs $k \rmv\in\rmv \tilde{\Set{K}}^{(j)}_{n}$.\\[1mm]
\hspace*{2mm} For all $j \rmv\in\rmv \{1,\ldots,J\}$, calculate estimates $\hat{p}\big(r_{k,n}^{(j)} \!\rmv=\! 1\big|\V{z}_{1:n}\big)$ of the existence probabilities 
  $p\big(r_{k,n}^{(j)} \!\rmv=\! 1\rmv\big|\V{z}_{1:n}\big)$, $k \rmv\in\rmv \tilde{\Set{K}}^{(j)}_{n}$ according to \eqref{eq:exist_prob_PFs} with $f\big({\V{a}}_{k,n}^{(j)}\ist, r_{k,n}^{(j)} \!\rmv=\! 1\big|\V{z}_{1:n}\big)$ replaced by $\tilde{q}\big(\V{a}_{k,n}^{(j)},1\big)$. \\[1mm]
\hspace*{2mm} For all $j \rmv\in\rmv \{1,\ldots,J\}$, determine the set $\Set{K}^{(j)}_n$ of PFs $k$ for which $\hat{p}\big(r_{k,n}^{(j)} \!\rmv=\! 1\rmv\big|\V{z}_{1:n}\big) \rmv>\rmv P_\text{prun}$.\\[1.7mm]
\textbf{\textit{Step 7---Detection and estimation}:}\\[.5mm]
\hspace*{2mm} Calculate an agent state estimate $\hat{\V{x}}_n$ according to \eqref{eq:mmse_agent} with $f(\V{x}_n |\V{z}_{1:n})$ replaced by $q(\V{x}_n)$. \\[.5mm]
\hspace*{2mm} For all $j \rmv\in\rmv \{1,\ldots,J\}$, determine the set $\hat{\Set{K}}^{(j)}_n$ of PFs $k \in \Set{K}^{(j)}_n$ for which $\hat{p}\big(r_{k,n}^{(j)} \!\rmv=\! 1\big|\V{z}_{1:n}\big) \rmv>\rmv P_\text{det}$, where $P_\text{det}$ is chosen larger than $P_\text{prun}$. \\[.5mm] 
\hspace*{2mm} For all $j \rmv\in\rmv \{1,\ldots,J\}$ and $k \rmv\in\rmv \hat{\Set{K}}^{(j)}_{n}\rmv$, calculate a PF state estimate $\hat{\V{a}}^{(j)}_{k,n}$ according to \eqref{eq:MMSE_PFs} 
  and \eqref{eq:marginal_PFs} with $f\big({\V{a}}_{k,n}^{(j)}\ist, r_{k,n}^{(j)} \!\rmv=\! 1\big|\V{z}_{1:n}\big)$ replaced by $\tilde{q}\big(\V{a}_{k,n}^{(j)},1\big)$.\\[1.7mm]
\textbf{\textit{Initialization}:}\\[.5mm]
\hspace*{2mm} The recursive algorithm is initialized at time $n\!=\!1$ by $ f(\V{x}_{1}) $ and $\Set{K}^{(j)}_{1} \!\rmv=\rmv \emptyset$, $j \rmv\in\rmv \{1,\ldots,J\}$.\\[-1mm]
\hrule
}

\section{Experimental Results}\label{sec:results}
To analyze the performance of the proposed BP-SLAM algorithm, we apply it to synthetic and real measurement data within two-dimensional (2-D) scenarios. The parameters involved in 
the algorithm and those used to generate the synthetic measurements are listed in Table~\ref{tab:simparam}.

\subsection{Analysis Setup}

\subsubsection{State-Evolution Model}\label{eq:stateEvoModel}
The agent's state-transition pdf $f(\V{x}_n|\V{x}_{n-1})$, with $\V{x}_n \rmv=\rmv  [\V{p}_n\trans \; \V{v}_n\trans ]\trans\rmv$, is defined by a linear, near constant-velocity motion model 
\cite[Sec. 6.3.2]{BarShalom2002EstimationTracking}, i.e., $\V{x}_n =  \V{A} \V{x}_{n-1} + \V{B} \V{w}_n$.
Here, $\V{A} \rmv\in\rmv \mathbb{R}^{4 \times 4}$ and $\V{B} \rmv\in\rmv \mathbb{R}^{4 \times 2}$ are as defined in \cite[Sec. 6.3.2]{BarShalom2002EstimationTracking} 
(with sampling period $\vu{\Delta T \rmv=\rmv 1\rmv}{s}$), and the driving process $\V{w}_n$ is iid across $\timestepsym$, zero-mean, and Gaussian with covariance matrix $\sigma_w^2\bold{I}_2$, where $\bold{I}_2$ denotes the $2 \rmv\times\rmv 2$ identity matrix. The PFs are static, i.e., the state-transition pdfs are given by $f\big(\tilde{\V{a}}_{k,n}^{(j)} \big| \V{a}_{k,n-1}^{(j)}\big) \rmv=\rmv \delta\big(\tilde{\V{a}}_{k,n}^{(j)}\rmv\!-\rmv \V{a}_{k,n-1}^{(j)}\big)$, where $\delta(\cdot)$ is the Dirac delta function. However, in our implementation of the BP-SLAM algorithm, we introduced a small driving process in the PF state-evolution model for the sake of numerical stability. Accordingly, the state evolution is modeled as $\tilde{\V{a}}_{k,n}^{(j)} \rmv\rmv=\rmv\rmv \V{a}_{k,n-1}^{(j)}\rmv+\rmv \V{\omega}_{k,n}^{(j)}$, where $\V{\omega}_{k,n}^{(j)}$ is iid across $k$, $n$, and $j$, zero-mean, and Gaussian with covariance matrix $\sigma_a^2\bold{I}_2$.

\begin{table*}[!t]
\caption{Simulation parameters.}
 \label{tab:simparam}
 \centering
 {\tabulinesep=0.8mm
 \begin{tabu}{
 |@{\hspace{.7mm}}c@{\hspace{.7mm}}	
 |@{\hspace{.7mm}}c@{\hspace{.7mm}}
 |@{\hspace{.7mm}}c@{\hspace{.7mm}}	
 |@{\hspace{.7mm}}c@{\hspace{.7mm}}
 |@{\hspace{.7mm}}c@{\hspace{.7mm}}
 |@{\hspace{.7mm}}c@{\hspace{.7mm}}
 |@{\hspace{.7mm}}c@{\hspace{.7mm}}
 |@{\hspace{.7mm}}c@{\hspace{.7mm}}
 |@{\hspace{.7mm}}c@{\hspace{.7mm}}
 |@{\hspace{.7mm}}c@{\hspace{.7mm}}
 |@{\hspace{.7mm}}c@{\hspace{.7mm}}
 |@{\hspace{.7mm}}c@{\hspace{.7mm}}
 |@{\hspace{.7mm}}c@{\hspace{.7mm}}|}
 	\hline \multicolumn{13}{|c|}{Parameters involved in the BP-SLAM algorithm} \\
   	\hline 
	$\sigma_w $ & $\sigma_a$ & $\sigma_{a,1}$ & $\sigma_m$ & $\mu_{\text{FA}}^{(j)}$ & $\mu_{\text{b}}^{(j)}$ & $\mu_{\text{n},1}^{(j)}$ & $P_\text{s}$ & $P_{\text{d}}$ & $P_\text{det}$ 
	& $P_\text{prun}$ & \#{\ist}particles & \#{\ist}simulation runs\\  
   	\hline 
	\vu{0.01}{$\text{m}/\text{s}^2$} & $10^{-4}${\ist}m, $0.5\rmv\rmv\cdot\rmv\rmv10^{-2}${\ist}m & $10^{-3}${\ist}m  & \vu{0.15}{$\text{m}$} & $1$, $2$ & $10^{-4}$ & $6$ & $0.999$ & $0.95$, $0.5$, $0.6$ & $0.5$ & $10^{-4}$ & $10^5$,  $3\rmv\rmv\cdot\rmv\rmv10^4$ & $100$, $30$ \\  \hline
 \end{tabu}}
 \begin{center}
 \vspace*{1mm}
 {\tabulinesep=0.9mm
 \begin{tabu}{|C{2cm}|C{2cm}|C{2cm}|}	
	\hline  \multicolumn{3}{|c|}{\hspace*{0mm}Parameters used to generate the synthetic measurements\hspace*{0mm}} \\
    \hline $\mu_{\text{FA}}^{(j)}$ & $P_{\text{d}}$ &$\sigma_{m,n}^{(j)}$ \\  
    \hline $1$, $2$ & $0.95$, $0.5$ & \vu{0.1\rmv}{$\text{m}$}  \\  \hline
 \end{tabu}}
 \end{center}
 \vspace*{-5mm} 
\end{table*}

\subsubsection{Measurement Model}
\label{sec:MeasModel}

In contrast to usual SLAM setups \cite{Thrun2005}, our measurement model is solely based on MPC ranges. The scalar range measurements $z_{m,n}^{(j)}$ are modeled as
\begin{equation}
	\label{eq:messmodel}
	z_{m,n}^{(j)} = \big\|\V{p}_n \!-\rmv \V{a}_{k,n}^{(j)} \big\| + \nu_{m,n}^{(j)} \ist,
\end{equation}
where the measurement noise $\nu_{m,n}^{(j)} $ is iid across $m$, $n$, and $j$, zero-mean, and Gaussian  with variance $\sigma_{m,n}^{(j)2}$. 
The measurement model \eqref{eq:messmodel} determines the likelihood function factors 
$f\big(\V{z}^{(j)}_{m,n} \big|\V{x}_n,\tilde{\V{a}}^{(j)}_{k,n} \big)$ and $f\big( \V{z}^{(j)}_{m,n} \big|\ist \V{x}_n, \breve{\V{a}}^{(j)}_{m,n} \big)$ in \eqref{eq:factorization_like_panchor}. 
However, we emphasize that the BP-SLAM algorithm can be extended to measurement models involving bearing measurements (AoAs and/or AoDs) or measurements from inertial measurement unit sensors. Such an extension would further increase the robustness of our approach.

\subsubsection{Common Simulation Parameters}\label{sec:sim_param}

The following parameters are used for both synthetic and real measurements, see also Table~\ref{tab:simparam}. 
We use the floor plan, agent trajectory, and two static PAs at positions $\V{a}_{1}^{(1)}$ and $\V{a}_{1}^{(2)}$ as shown in Fig.~\ref{fig:VAgeometry_SLAM}.
The false alarm pdf $f_{\text{FA}}\big( z_{m,n}^{(j)} \big)$ is uniform on $[0\ist\text{m},30\ist\text{m}]$ with mean number $\mu_\text{FA}^{(j)}$. 
The conditional pdf $f_{\text{n},n}\big(\breve{\V{a}}^{(j)}_{m,n}\big|\V{x}_n\big)$ and mean number $\mu_{\text{n},n}^{(j)}$ of newly detected features are inferred online 
by a PHD filter using a birth pdf $f_{\text{b},n}\big(\breve{\V{a}}^{(j)}_{m,n}\big|\V{x}_n\big)$ and a mean number of newborn features $\mu_\text{b}^{(j)}$ 
(see \cite{Horridge2011PHD}). The birth pdf is uniform on the region of interest (ROI), which is a circular disk of radius 30{\ist}m around the center of the floor plan shown in Fig.~\ref{fig:VAgeometry_SLAM}.  
At time $n \!=\! 1$, newly detected features are initialized by setting $\mu_{\text{n}, 1}^{(j)} \rmv=\rmv 6$ and using an initial pdf $f_{\text{n},1}\big(\breve{\V{a}}^{(j)}_{m,1}\big|\V{x}_1\big)$ that is uniform on the ROI. The detection probability is a constant value for all PFs and PAs, i.e. $P^{(j)}_\text{d}\big(\V{x}_n,\V{a}^{(j)}_{k,n}\big)=\rmv P_\text{d}$.
The values of these and further parameters are given in Table~\ref{tab:simparam}.

Our implementation of the BP-SLAM algorithm uses a particle representation of messages and
beliefs similarly to \cite{MeyerTSP2017,MeyerTSPIN2016}. The particles for the initial PA states are drawn from the 2-D Gaussian distributions $\mathcal{N}\big(\V{a}_{1,1}^{(j)}, \sigma_{a,1}^2 \bold{I}_2\big)$, where  $\V{a}_{1,1}^{(j)}$ is the position of PA $j \!\in\! \{1,2\}$ and $\sigma_{a,1} \rmv= 10^{-3}\ist\text{m}$. This implies that the initial PA positions are effectively known. On the other hand, we do not use any prior information about the VA states. The particles for the initial agent state are drawn from a 4-D uniform distribution with center $\V{x}_1 = [\V{p}_{1}\trans\;0\;\, 0]\trans\rmv$, where $\V{p}_{1}$ is the starting position of the actual agent trajectory, and the support of each component about the respective center is given by $[-\lambda, \lambda]$. Here, $\lambda$ is $0.5$ except for our simulations in Section \ref{sec:FastSLAM}, where it is $0.1$; the physical dimension of $\lambda$ is m (position) or m/s (velocity).
We note that the BP-SLAM algorithm performs well even without any prior information about the initial states of the mobile agent and the PAs. However, because we use only range measurements, the estimated feature map and agent trajectory would contain an arbitrary translation and rotation relative to the true positions. Finally, the number $P$ of message passing iterations for DA is limited by the termination condition $\big[\sum_{k\in\Set{K}_{n-1}^{(j)}} \sum_{m \in \Set{M}_n^{(j)}} \!\big(\nu_{m\rightarrow k}^{(p)}(c^{(j)}_{k,n}) - \nu_{m\rightarrow k}^{(p-1)}(c^{(j)}_{k,n})\big)^2\big]^{1/2} \!\rmv< 10^{-7}$ (cf.\ \eqref{eq:featureDARV}) or by the maximum number $P_\text{max} \!=\! 1000$.

\subsection{Results for Synthetic Measurements}
\label{sec:sim_param:syn}

For our simulations based on synthetic measurements, we used the common simulation parameters described above. The range measurements were generated with noise standard deviation $\sigma_{m,n}^{(j)} \!=\rmv 0.1${\ist}m. However, for numerical robustness, the BP-SLAM algorithm used noise standard deviation $\sigma_m \!=\! 1.5\cdot\sigma_{m,n}^{(j)}$. The standard deviation of the driving process in the PF state-evolution model
was \vu{\sigma_a \rmv= 10^{-4}}{m}. We performed 100 simulation runs, each using the floor plan and agent trajectory shown in Fig.~\ref{fig:VAgeometry_SLAM}. In each run, we generated with detection probability $P_\text{d}$ noisy ranges $z_{m,n}^{(j)}$ according to \eqref{eq:messmodel}. 
Evaluation of \eqref{eq:messmodel} was based on the agent positions along the agent trajectory, the fixed PA positions $\V{a}_{1}^{(1)}$, $\V{a}_{1}^{(2)}$, and the  fixed positions of the first-order VAs, i.e., $\V{a}_{l}^{(j)} \!\rmv\in\rmv \mathbb{R}^2\rmv$, $l \rmv=\rmv 2,\ldots,L_n^{(j)}$ for $j \!=\! 1,2$ (all shown in Fig.~\ref{fig:VAgeometry_SLAM}), where $L_n^{(1)} \!=\rmv 6$ and $L_n^{(2)} \!=\rmv 5$ are the numbers of features (PA plus associated VAs). In addition, false alarm measurements $z_{m,n}^{(j)}$ were generated as described below.

\subsubsection{Comparison of Different Parameter Settings}
\label{sec:sim_setups} 

\begin{figure*}[t!]
\centering
\includegraphics[scale=1]{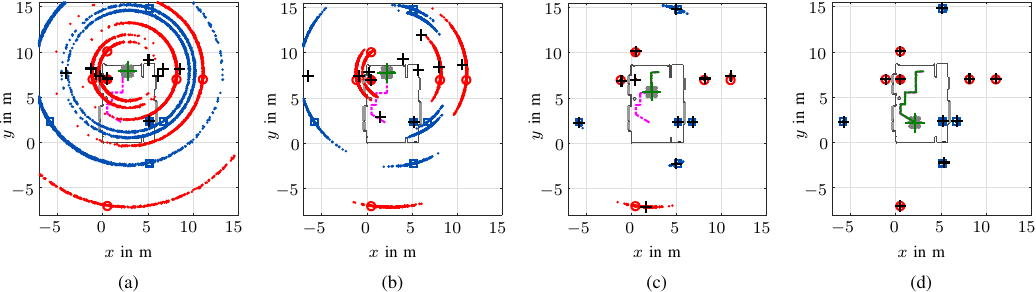}
\vspace*{2mm}
   \caption{
 Particle convergence. Particles representing the posterior pdfs of the state of the mobile agent (gray) and of the states of the detected PFs (red for PA 1, blue for PA 2) are shown at 
 (a) $n \!=\! 30$, (b) $n \!=\! 90$, (c) $n \!=\! 300$, and (d) $n \!=\! 900$. The positions of PA 1 and PA 2 are indicated by a red bullet and a blue box, respectively, and the corresponding geometrically expected VA positions by red circles and blue squares. The green line represents the MMSE estimates of the past mobile agent positions and the green cross the MMSE estimate of the current mobile agent position. The black crosses indicate the MMSE estimates of the positions of the detected PFs.}                                                                         
  \label{fig:syntheticdata_PFconv} 
\vspace*{4mm}
\end{figure*}

\begin{figure*}[t!]
\centering
\includegraphics[scale=1]{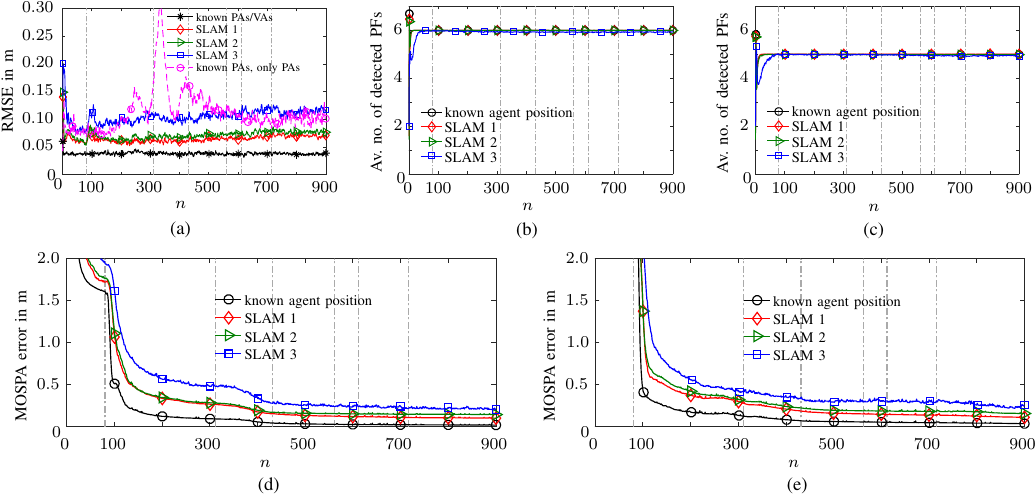}
\vspace*{1mm}
\caption{Results for synthetic measurements: (a) Agent position RMSE, 
(b) and (c) average number of detected PFs associated with PA 1 and 2, respectively, 
(d) MOSPA error for PA 1 and the associated VAs, and (e) MOSPA error for PA 2 and the associated VAs. 
In (a), the curve labeled ``known PAs/VAs'' shows the comparison to the benchmark algorithm that assumes knowledge of the feature map, i.e., of the PA and VA positions, and the curve labeled ``known PAs, only PAs'' shows the comparison to the benchmark algorithm using only the two PA positions, i.e., no VA positions. In (b)--(e), the curves labeled ``known agent position'' show the case where the agent position is known. 
In (b) and (c), the curves labeled ``known agent position,'' ``SLAM 1,'' and ``SLAM 2'' coincide. 
The vertical dash-dotted lines indicate the times around which the mobile agent performs turns.}
\label{fig:syntheticdata_agenterrors_MOSPA_numfeat}
\vspace*{-3mm}
\end{figure*}

We considered three different parameter settings dubbed SLAM 1, SLAM 2, and SLAM 3. In SLAM 1 and SLAM 2, we used detection probability $\rmv P_\text{d} \rmv=\rmv 0.95$ and mean number of false alarms $\mu_\text{FA}^{(j)} \!=\! 1$. The posterior pdfs of the agent state, of the legacy PF states, and of the new PF states were each represented by 100.000 (SLAM 1) or 30.000 (SLAM 2) 
particles. (The number of particles could be reduced if, e.g., the MPCs' AoAs were used in addition to the range measurements since this would decrease
the effective support regions of the posterior pdfs, or by adaptively adjusting the detection probability $P_\text{d}$ as in \cite{LeitingerICC2017}.)
In SLAM 3, we used $P_\text{d} \rmv=\rmv 0.5$ and $\mu_\text{FA}^{(j)} \!\rmv=\! 2$ to analyze the robustness of the BP-SLAM algorithm to extremely poor radio signal conditions, i.e., to scenarios characterized by a high probability that existing MPCs are not detected or nonexisting MPCs are detected; furthermore, the agent state and the PF states were each represented by 100.000 particles. 
For one exemplary simulation run, Fig.~\ref{fig:syntheticdata_PFconv} illustrates the convergence of the posterior pdfs of the PF positions to the true feature (PA and VA) positions by displaying the respective particles at times $n \!=\! 30, 90, 300, 900$. These results demonstrate that the BP-SLAM algorithm is able to cope with highly multimodal distributions and with measurements conveying only very limited information at each time step (since only range measurements are used).

Fig.~\ref{fig:syntheticdata_agenterrors_MOSPA_numfeat} shows the root mean square error (RMSE) of the estimated time-varying agent position, the average numbers of detected PFs for the two PAs, and the mean optimal subpattern assignment (MOSPA)\footnote{We 
remark that an alternative to the ``classical'' OSPA metric \cite{Schuhmacher2008} is provided by the generalized OSPA metric proposed in \cite{RahmathullahFusion2017}, which does not normalize the OSPA error by the cardinality of the larger set and penalizes the cardinality error 
differently.} 
errors \cite{Schuhmacher2008} for the two PAs and the associated VAs, all versus time $n$. These results were obtained by averaging over the 100 simulation runs. The MOSPA errors are based on the Euclidean metric and use cutoff parameter 5{\ist}m and order 1 \cite{Schuhmacher2008}. It can be seen in Fig.~\ref{fig:syntheticdata_agenterrors_MOSPA_numfeat}(a) that the agent position RMSE is mostly below 0.12{\ist}m for all parameter settings and below 0.08{\ist}m for SLAM 1 and SLAM 2. The average numbers of detected PFs in Figs.~\ref{fig:syntheticdata_agenterrors_MOSPA_numfeat}(b), (c) are effectively equal to the respective true numbers of features $L_n^{(1)} \!\rmv=\rmv 6$ and $L_n^{(2)} \!\rmv=\rmv 5$ for SLAM 1 and SLAM 2. The MOSPA errors are shown in Figs.~\ref{fig:syntheticdata_agenterrors_MOSPA_numfeat}(d), (e). For SLAM 1, they decrease until they are ultimately below 0.11{\ist}m for both PAs. In general, the agent position RMSE and MOSPA errors for SLAM 2 and SLAM 3 are slightly larger. However, the 
performance of our algorithm is still good, which suggests a high robustness. We note that the agent position RMSE converged for all simulation runs and all parameter settings. Furthermore, we observed that the position RMSEs of the individual PFs (not shown in Fig.~\ref{fig:syntheticdata_agenterrors_MOSPA_numfeat}) are at most 0.25{\ist}m and in many cases around 0.05{\ist}m.

As a performance benchmark for the accuracy of agent localization, we also plot in Fig.~\ref{fig:syntheticdata_agenterrors_MOSPA_numfeat}(a) the agent position RMSE
obtained for SLAM 1 with the algorithm from \cite{LeitingerGNSS2016} (referred to as ``benchmark algorithm''). This algorithm uses BP-based probabilistic DA 
just as our BP-SLAM algorithm but assumes knowledge of the feature map, i.e., of the PA and VA positions. It is seen that its agent position RMSE (labeled ``known PAs/VAs'' in Fig.~\ref{fig:syntheticdata_agenterrors_MOSPA_numfeat}(a)) is significantly lower than that of the BP-SLAM algorithm,
which  demonstrates the impact of map uncertainty on the performance of SLAM. In addition, Fig.~\ref{fig:syntheticdata_agenterrors_MOSPA_numfeat}(a) shows the agent position RMSE
obtained for SLAM 1 with the benchmark algorithm using only the two PA positions, i.e., no VA positions (labeled ``known PAs, only PAs'' in Fig.~\ref{fig:syntheticdata_agenterrors_MOSPA_numfeat}(a)). Here, the measurement vector $\V{z}_n^{(j)}$ was preprocessed such that only the measurement with the shortest range was detected and used by the algorithm. Since for the chosen PA positions the mobile agent position cannot be estimated unambiguously, it is not possible to determine the agent position RMSE without a significant bias. Therefore, to be able to determine the agent position RMSE along the true trajectory, we applied genie-aided k-means clustering of the particles. One can see that the agent position RMSE of the benchmark algorithm
now tends to be considerably higher, especially around the second turn of the trajectory where the agent crosses the line between the PAs (cf.\ Fig.~\ref{fig:VAgeometry_SLAM}).

In addition, as a performance benchmark for the accuracy of feature map estimation, we compare in Figs.~\ref{fig:syntheticdata_agenterrors_MOSPA_numfeat}(b)--(e) the average number of detected PFs and the MOSPA errors to those that would be obtained for the SLAM 1 parameter setting if the agent position was known at all times. These two benchmarks provide bounds on the two main performance aspects of SLAM, i.e., accuracy in localization and mapping.

Finally, we considered the use of measurement gating, which is commonly employed in practical implementations to reduce computational complexity \cite[Sec.\ 2.3.2]{BarShalom11}.
With measurement gating, DA is performed only on those measurements that fall into given ``gates'' around predicted measurement values. We chose the gating threshold as $\gamma \!=\! 6.635$, which implies that the probability that feature-originated measurements are outside their corresponding gate is $10^{-2}$ \cite[Table 2.3.2-1]{BarShalom11}. For SLAM 2, we measured the average runtime per time step $n$, averaged over the 900 time steps and 100 simulation runs, as 0.0929{\ist}s without measurement gating and 0.0735{\ist}s with measurement gating. Thus, measurement gating yielded a reduction of the average runtime by 20.1{\ist}\%, and this did not come at the cost of a noticeable performance loss in terms of agent position RMSE or MOSPA error. These results were obtained using a MATLAB implementation on an Intel i5-4690 CPU. We note that besides measurement gating, another potential means of reducing complexity is the use of geometric data structures such as kd-trees \cite{Moore1998VeryFE}. For example, kd-
trees were used in the RB-SLAM (FastSLAM) algorithm
reported in \cite{ MontemerloAAAI2002}. However, in the particle-based implementation of the proposed BP-SLAM algorithm, due to the necessity of performing probabilistic DA,
the weights representing the agent state and the feature states are updated by a weighted sum of measurements instead of a single measurement (as is done in established particle-based RB-SLAM algorithms, even when Monte Carlo-based DA is used \cite{MontemerloAAAI2002, Thrun2005}). Due to this difference, to the best of our knowledge, geometric data structures such as kd-trees cannot be directly applied to SLAM with probabilistic DA.

\subsubsection{Comparison with Rao-Blackwellized SLAM}
\label{sec:FastSLAM}

\begin{figure*}[t!] 
\includegraphics[scale=1]{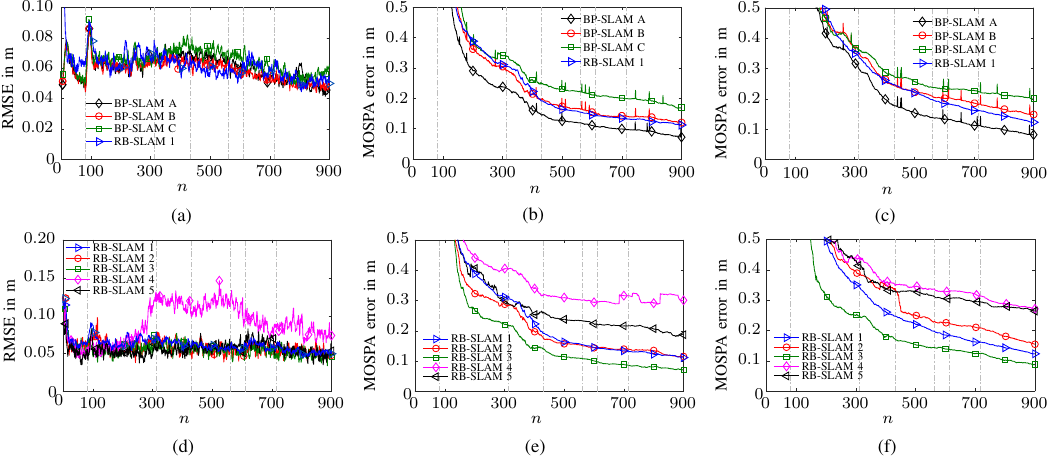}
\vspace*{2mm}
\caption{Results for synthetic measurements; comparison between BP-SLAM and RB-SLAM \cite{DurrantWhyte2006,MontemerloAAAI2002,GentnerTWC2016}: 
(a) Agent position RMSE, 
(b) MOSPA error for PA 1 and the associated VAs, and (c) MOSPA error for PA 2 and the associated VAs, all for BP-SLAM A--C and RB-SLAM 1;
(d)--(f) comparison of agent position RMSE and PA/VA MOSPA error results for RB-SLAM 1--5. 
The vertical dash-dotted lines indicate the times around which the mobile agent performs turns.}
\label{fig:syntheticdata_agenterrors_MOSPA_RB_BP}
\end{figure*}

Next, we compare the accuracy and complexity (runtime) of our BP-SLAM algorithm with those of RB-SLAM \cite{DurrantWhyte2006,MontemerloAAAI2002,GentnerTWC2016}, which is an important state-of-the-art method. We note that RB-SLAM can be motivated by the idea of stacking the agent state and all the feature states into one high-dimensional joint state and using a particle filter to track the joint state. A direct implementation of this idea is typically infeasible due to the curse of dimensionality \cite{Thrun2005}, which is further exacerbated by our range-only measurement model. Indeed, even the 100.000 particles used in simulation setup SLAM 1 
would be totally insufficient to represent the joint state vector. The RB-SLAM algorithm avoids the curse of dimensionality by exploiting the conditional independencies of the feature states (conditionally on the agent state) and applying a Rao-Blackwellization to the joint state \cite[Sec. 13]{Thrun2005}, \cite{MontemerloAAAI2002}. 

Because our measurement model is highly nonlinear, we adopt the RB-SLAM implementation proposed in \cite{GentnerTWC2016}, which employs particle filters instead of EKFs to estimate the feature states. (Note that we only implemented the SLAM algorithm proposed in \cite{GentnerTWC2016}, not the full two-stage method that includes a channel estimator/tracker.)
Since the measurement-feature association is unknown, we perform Monte Carlo-based DA for each particle separately,
as it is commonly done in classical RB-SLAM \cite{DurrantWhyte2006,MontemerloAAAI2002}. The classical RB-SLAM algorithm is very sensitive to missed detections and clutter measurements, especially for our range-only measurement model, which implies a strong DA uncertainty. Therefore, to avoid the risk of divergence in RB-SLAM, we assumed in our simulation that there are no missed detections or false alarms, i.e., we generated the data using $\rmv P_\text{d} \rmv=\! 1$ and $\mu_\text{FA}^{(j)} \!=\! 0$.

Fig.~\ref{fig:syntheticdata_agenterrors_MOSPA_RB_BP} shows the RMSE of the estimated agent position and the MOSPA errors for the two PAs and the associated VAs.
For BP-SLAM, we considered three different parameter settings, dubbed BP-SLAM A--C, in which each posterior state pdf (for the agent state, each legacy PF state, and each new PF state) was represented by 50.000 (BP-SLAM A), 10.000 (BP-SLAM B), or 5.000 (BP-SLAM C) particles. 
For RB-SLAM, we considered five different parameter settings, dubbed RB-SLAM 1--5, in which the posterior pdf of the agent state and the posterior pdf of each feature state
were represented by, respectively,
50 and 10.000 (RB-SLAM 1),
100 and 5.000 (RB-SLAM 2),
100 and 10.000 (RB-SLAM 3),
500 and 1.000 (RB-SLAM 4),
or 1.000 and 1.000 (RB-SLAM 5) particles.
In Fig.~\ref{fig:syntheticdata_agenterrors_MOSPA_RB_BP}(a), it can be seen that the agent position RMSEs obtained with BP-SLAM A--C and with RB-SLAM 1 are generally quite similar. 
On the other hand, Figs.~\ref{fig:syntheticdata_agenterrors_MOSPA_RB_BP}(b) and (c) show that, as may be expected, the MOSPA error of BP-SLAM is smaller for a larger number of particles. 
The MOSPA error of RB-SLAM~1 is seen to be only slightly smaller than that of BP-SLAM B 
(which is a good basis for a comparison, because also RB-SLAM 1 uses 10.000 particles for each feature state). 
This is remarkable, in view of the fact that BP-SLAM assumes a priori independence of the agent state and all the feature states. 
BP-SLAM A even has a significantly smaller MOSPA error than RB-SLAM. Figs.~\ref{fig:syntheticdata_agenterrors_MOSPA_RB_BP}(d)--(f) compare the RMSE and MOSPA error results for the different RB-SLAM parameter settings; these results will be discussed presently. 
\begin{table}[!t]
\caption{Average runtimes of BP-SLAM A--C and RB-SLAM 1--5 per time step.}
\vspace*{-1mm}
\label{tab:runtimes}
\begin{center}
{\tabulinesep=0.8mm
 \begin{tabu}{|C{1.7cm}|C{1.7cm}|C{1.7cm}|C{1.7cm}|}	
    \hline BP-SLAM A & BP-SLAM B & BP-SLAM C  \\ \hline 
    \vu{0.146\rmv}{s}  & \vu{0.037\rmv}{s}  & \vu{0.028\rmv}{s}  \\  \hline
 \end{tabu}}
 \end{center} 
 \begin{center}
 {\tabulinesep=0.9mm
\begin{tabu}{
 |@{\hspace{.8mm}}c@{\hspace{.8mm}}
 |@{\hspace{.8mm}}c@{\hspace{.8mm}}
 |@{\hspace{.8mm}}c@{\hspace{.8mm}}
 |@{\hspace{.8mm}}c@{\hspace{.8mm}}
 |@{\hspace{.8mm}}c@{\hspace{.8mm}}|
}	
\hline RB-SLAM 1 & RB-SLAM 2 & RB-SLAM 3 & RB-SLAM 4  & RB-SLAM 5 \\ \hline 
\vu{0.47\rmv}{s} & \vu{0.62\rmv}{s}  & \vu{1.26\rmv}{s}  & \vu{0.65\rmv}{s} & \vu{1.61\rmv}{s} \\  \hline
 \end{tabu}}
 \end{center}  
 \vspace*{-4mm}
\end{table}

Table~\ref{tab:runtimes} shows the average runtimes per time step $n$ (averaged over 900 time steps and 100 simulation runs) of our MATLAB implementations of the algorithms on an Intel i5-4690 CPU. It is seen that BP-SLAM is significantly less complex than RB-SLAM 1; in particular, the runtime of BP-SLAM B---i.e., using the same number of particles as RB-SLAM 1, leading to only slightly poorer performance as observed above---is smaller by a factor of more than 10. 
This is because RB-SLAM calculates the distances of each agent state particle to each feature state particle. Consequently, the complexity of RB-SLAM 
scales linearly in the \emph{product} of the number of agent particles and the total number of feature particles; by contrast, the complexity of BP-SLAM scales linearly 
in the number of all particles. One can conclude that the slightly better accuracy of RB-SLAM relative to BP-SLAM B comes at the expense of a much highercomplexity. We note that the complexity of RB-SLAM algorithms that use particles to represent the feature states can be reduced by the adaptive resampling algorithm
presented in \cite{GentnerHindawi2017}.

From these runtime results and from Figs.~\ref{fig:syntheticdata_agenterrors_MOSPA_RB_BP}(d)--(f), the following further conclusions can be drawn: 
(i) Increasing the number of agent particles in RB-SLAM increases the accuracy of agent state and VA position estimation only slightly but increases the computational complexity significantly. 
(ii) Decreasing the number of VA particles in RB-SLAM decreases the accuracy of agent state  and VA position estimation significantly and may even
lead to a divergence of the agent state since the VA particles converge to wrong positions.  
(Here, we consider a simulation run to be divergent if the error rises above 30 cm and continues to increase. Note that the agent position RMSE and 
the PA/VA MOSPA error plotted in Figs.~\ref{fig:syntheticdata_agenterrors_MOSPA_RB_BP}(d)--(f) were calculated using only the converged simulation runs for the respective parameter setting. For parameter settings RB-SLAM 1--3, all simulation runs converged. For parameter settings RB-SLAM 4 and RB-SLAM~5, only 73\,\% and 87\,\%, respectively, of the simulation runs converged.)

\subsection{Results for Real Measurements}
\label{sec:sim_param:real}

For an evaluation of the performance of the proposed BP-SLAM algorithm using real measurements, we chose $P^{(j)}_\text{d}\big(\V{x}_n,\V{a}^{(j)}_{k,n}\big)=\rmv P_\text{d} \rmv=\rmv 0.6$ and $\mu_\text{FA}^{(j)} \rmv=\rmv 2$. 
This accounts for the lower detection probability and higher false alarm probability exhibited by the preliminary channel estimation due to the diffuse multipath existing in indoor environments.

The measurements were taken from the seminar room scenario previously used in \cite{MeissnerWCL2014, LeitingerICC2017}. 
They correspond to five closely spaced parallel trajectories each consisting of 900 agent positions with a spacing of 0.01{\ist}m along each trajectory and a spacing of 0.01{\ist}m between the trajectories, resulting in a total number of 4500 agent positions. The magenta line in Fig.~\ref{fig:VAgeometry_SLAM} represents one of the five trajectories. More details about the measurements are provided in \cite{MeissnerWCL2014}; in particular, a close-up of the five trajectories is shown in \cite[Fig.~1]{MeissnerWCL2014}. At each agent position, the agent transmitted an ultra-wideband signal, which was received by the two static PAs. 
This signal was measured using an M-sequence correlative channel sounder with frequency range 3--10{\,}GHz and antennas with an approximately uniform radiation pattern in the azimuth plane and zeros in the floor and ceiling directions. Within the measured band, the actual signal band was selected by a filter with raised-cosine impulse response $s(t)$ with a roll-off 
factor of 0.5, a two-sided 3-dB bandwidth of 2{\,}GHz, and a center frequency of 7{\,}GHz. From the measured signals, the range measurements $z_{m,n}^{(j)} \rmv= c \ist \hat{\tau}_{m,n}^{(j)}$ constituting the input to the proposed algorithm were derived by means of a snapshot-based SISO SAGE algorithm \cite{Fleury1999} that provides estimates $\hat{\tau}_{m,n}^{(j)}$ of the delays of the MPCs and estimates of the associated complex amplitudes (cf. \eqref{eq:rx_signal}). 
In this method, the number of estimated MPCs for each PA $j$ was fixed to $M_{n}^{(j)} \!\rmv=\rmv\rmv 20$. Estimates of the range variances $\sigma_{m,n}^{(j)2}$ (cf.\ \eqref{eq:messmodel}) were determined from the estimated complex amplitudes as described in \cite{MeissnerWCL2014, LeitingerICC2017}. The standard deviation of the driving process in the PF state-evolution model was chosen as \vu{\sigma_a \rmv= 0.5\cdot 10^{-2}}{m}. We performed 30 simulation runs. The pdfs of the states were represented by 30.000 particles each. Each individual trajectory was processed independently, i.e., 
the estimated PF positions of a trajectory were not used as prior knowledge for another trajectory. 

\begin{figure}[t!]
 \centering	
 \includegraphics[scale=1]{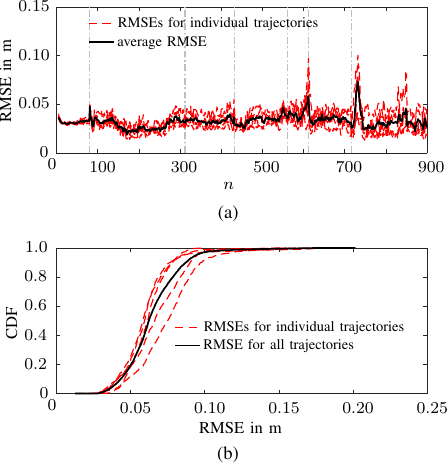}
 \vspace*{1mm}
    \caption{Results for real measurements: (a) Agent position RMSEs for the five individual trajectories and average RMSE (averaged over the five trajectories), (b) empirical CDFs of the RMSEs for the five individual trajectories and empirical CDF of 
  the five RMSEs taken together.}
\label{fig:realdata_agenterrors_cdf}
\vspace*{-3mm}
\end{figure}

Fig.~\ref{fig:realdata_agenterrors_cdf}(a) shows the agent position RMSEs of BP-SLAM obtained individually for the five trajectories versus time $n$, 
along with the overall RMSE averaged over the five trajectories. Fig.~\ref{fig:realdata_agenterrors_cdf}(b) shows the empirical cumulative distribution function (CDF) of the individual RMSEs and the empirical CDF of 
the five RMSEs taken together. It can be seen that the individual CDFs are very close to 1 already at RMSE equal to 0.12{\ist}m or even less. The maximum 
of all the individual RMSEs is below 0.2{\ist}m in all cases and below 0.083{\ist}m in 90\% of all cases.\footnote{In \cite{LeitingerICC2017}, lower RMSE values (below 0.072{\ist}m in all cases and below 0.035{\ist}m in 90\% of all cases) were obtained by using an adaptive adjustment of the detection probability 
$P_\text{d}$ in the SLAM algorithm.} 

\begin{figure}[t!]
\centering
\includegraphics[width=0.65\columnwidth]{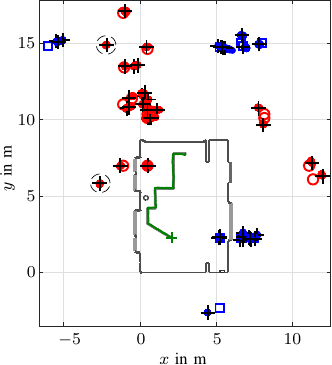}
 \vspace*{2mm}
 \caption{Results for real measurements: Particles representing the posterior pdfs of the states of the detected PFs (red for PA 1, blue for PA 2). The positions of PA 1 and PA 2 are indicated by a red bullet and a blue box, respectively, and the corresponding geometrically expected VA positions by red circles and blue squares. The green line represents the MMSE estimates of the mobile agent positions, the last of which is indicated by a green cross. The black crosses indicate the MMSE estimates of the positions of the detected PFs.
 The dashed circles indicate PF positions that cannot be associated with geometrically expected VA positions.}                                                                    
\label{fig:realdata_PFconv} 
\vspace*{-1mm}
\end{figure}

For an exemplary simulation run, Fig.~\ref{fig:realdata_PFconv} depicts the particles representing the posterior pdfs of the states of the detected PFs as well as the MMSE estimates of the positions of the detected PFs at the current time instant for the five trajectories. Furthermore, Fig.~\ref{fig:realdata_PFconv} depicts the MMSE estimates of the past and current mobile agent positions. Almost all the estimated PF positions can be associated with geometrically expected VA positions. This shows that the BP-SLAM algorithm is able to leverage position-related information contained in the radio signals for accurate and robust localization. 

\section{Conclusions and Future Perspectives}\label{sec:concl}
We proposed a radio signal based SLAM algorithm with probabilistic DA. The underlying system model describes specular MPCs in terms of VAs with unknown and possibly time-varying positions. To tackle the DA problem, i.e., the unknown association of MPCs with VAs, we modeled the entire SLAM problem including probabilistic DA in a Bayesian framework. We then represented the factorization of the joint posterior distribution by a factor graph and applied BP for approximate marginalization of the joint posterior distribution. This approach allowed the incorporation of an efficient BP algorithm for probabilistic DA that was originally proposed for MTT \cite{MeyerTSP2017, WilliamsTAE2014}. Our factor graph extends that of \cite{LeitingerICC2017} by the states of new potential features. 

Simulation results using synthetic data showed that the proposed BP-SLAM algorithm estimates the time-varying agent position and the feature map with high accuracy and robustness, even in conditions of strong clutter and low probability of detection. Moreover, an experimental analysis using real ultra-wideband radio signals in an indoor environment showed that the BP-SLAM algorithm performs similarly well in real-world scenarios; the agent position error was observed to be below 0.2{\ist}m for 100\% and below 0.083{\ist}m for 90\% of all measurements.

Promising directions for future research are to exploit further MPC parameters such as AoAs and AoDs, to include additional types of features such as scatter points, and to redefine the features to be extended objects. Finally, studying operation in an unsynchronized sensor network and a distributed (decentralized) mode of operation would be interesting.


\bibliographystyle{IEEEtran}
\bibliography{references}


\begin{IEEEbiography}[{\includegraphics[width=25mm,height=32.15mm,clip,keepaspectratio]{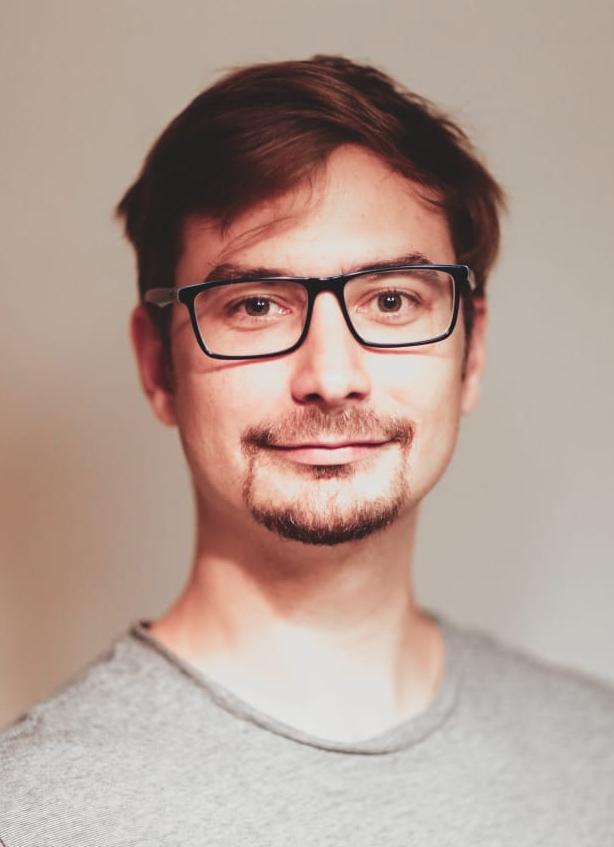}}]{Erik~Leitinger} (S'12--M'16) received his Dipl.-Ing.\ (M.Sc.\ ) and Ph.D.\ degrees (with highest honors) in electrical engineering from Graz University of Technology, Austria in 2012 and 2016, respectively. He is currently a Postdoctoral Associate at Graz University of Technology. He was Postdoctoral Associate at the department of Electrical and Information Technology at Lund University from 2016 to 2018. He is an Erwin Schr\"odinger Fellow.

Erik Leitinger received the outstanding award of excellence of the Federal Ministry of Science, Research and Economy (BMWFW) for his Ph.D.\ Thesis. Dr.\ Leitinger served as reviewer for several IEEE journals and as a TPC member in IEEE conferences. His research interests include localization and navigation, stochastic modeling and estimation of radio channels, factor graphs and iterative message passing algorithms, and estimation/detection theory.
\end{IEEEbiography}

\begin{IEEEbiography}[{\includegraphics[width=25mm,height=32.15mm,clip,keepaspectratio]{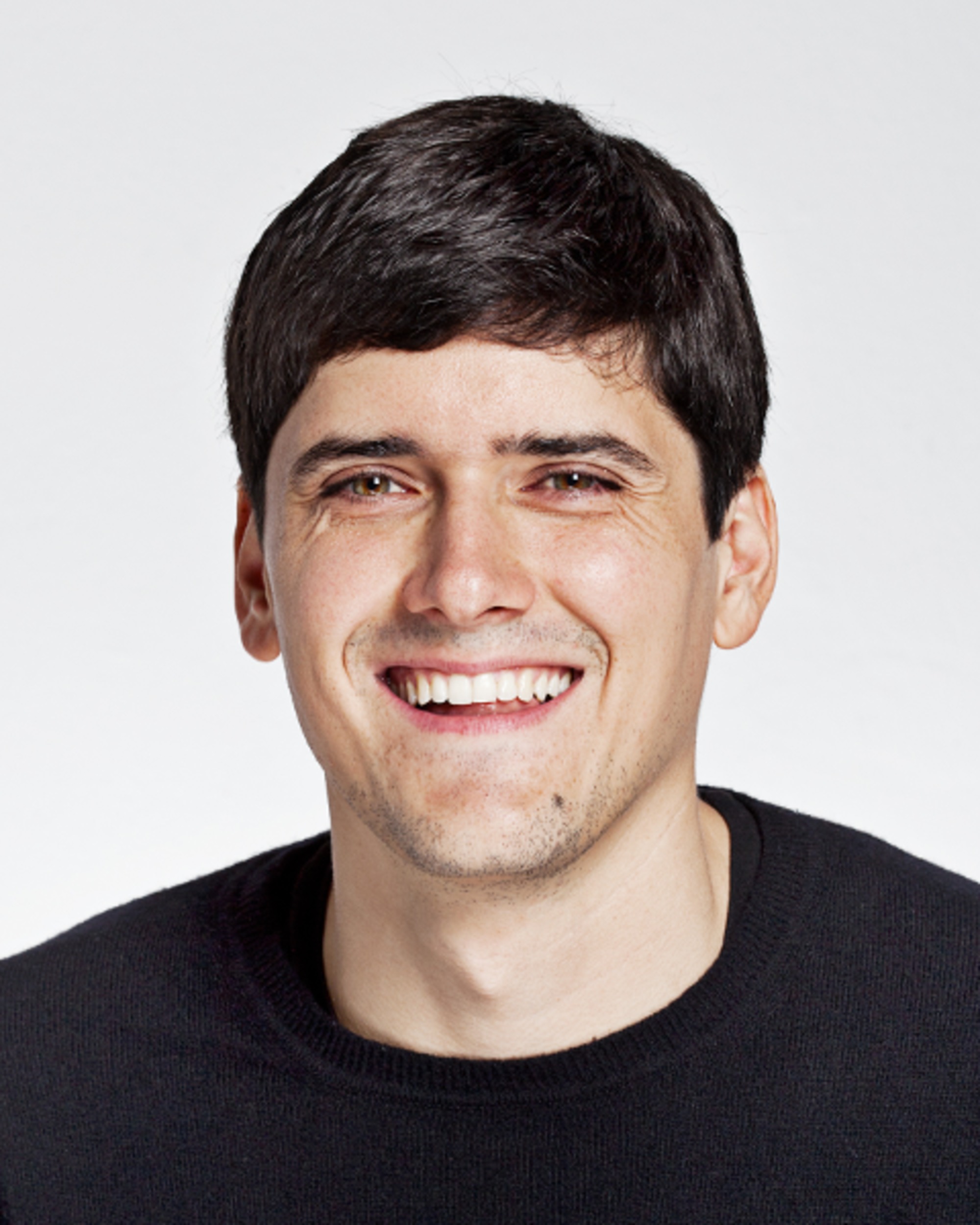}}]{Florian~Meyer} (S'12--M'15) received the Dipl.-Ing.\ (M.Sc.\ ) and Ph.D.\ degrees in electrical engineering from TU Wien, Vienna, Austria in 2011 and 2015, respectively. He is currently a postdoctoral researcher with the Laboratory for Information and Decision Systems, Massachusetts Institute of Technology (MIT). He was a visiting researcher with the Department of Signals and Systems, Chalmers University of Technology, Gothenburg, Sweden in 2013 and with the NATO Centre for Maritime Research and Experimentation (CMRE), La Spezia, Italy in 2014 and 2015. In 2016, he joined CMRE as a research scientist. Dr.\ Meyer is an Associate Editor of the Journal of Advances in Information Fusion and served as a co-chair of the IEEE ANLN Workshop at IEEE ICC 2018, Kansas City, MO, USA and at IEEE ICC 2019, Shanghai, China. His research interests include multiobject tracking, applied ocean sciences, 
network localization and navigation, and multiagent systems. He is an Erwin Schr\"odinger Fellow.
\end{IEEEbiography}

\begin{IEEEbiography}[{\includegraphics[width=25mm,height=32.15mm,clip,keepaspectratio]{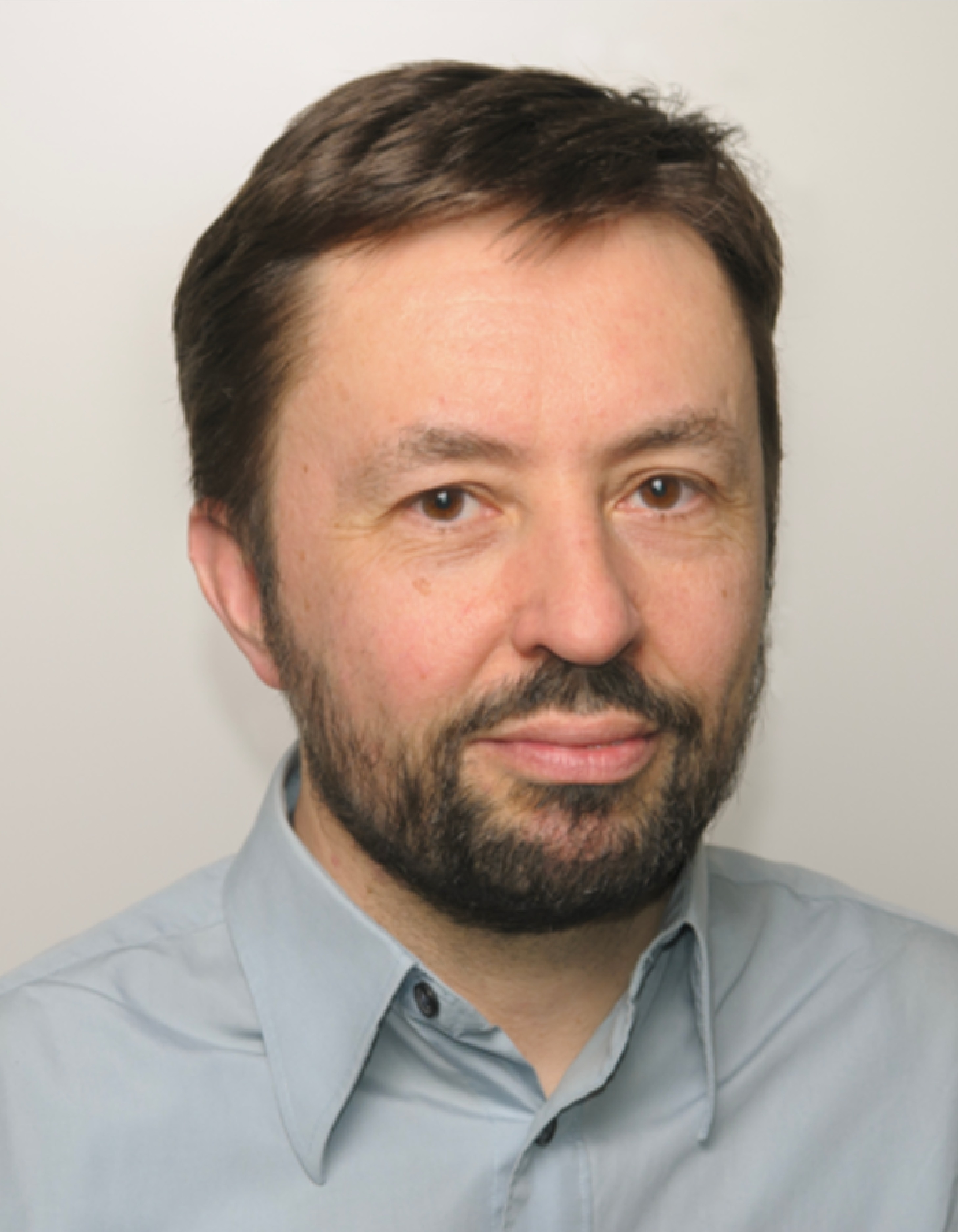}}]{Franz~Hlawatsch}(S'85--M'88--SM'00--F'12) received the Diplom-Ingenieur, Dr. techn.\, and Univ.-Dozent (habilitation) degrees in electrical engineering/signal processing from TU Wien, Vienna, Austria in 1983, 1988, and 1996, respectively. Since 1983, he has been with the Institute of Telecommunications, TU Wien, where he is currently an Associate Professor. During 1991--1992, as a recipient of an Erwin Schr\"odinger Fellowship, he spent a sabbatical year with the Department of Electrical Engineering, University of Rhode Island, Kingston, RI, USA. In 1999, 2000, and 2001, he held one-month Visiting Professor positions with INP/ENSEEIHT, Toulouse, France and IRCCyN, Nantes, France. He (co)authored a book, three review papers that appeared in the {\sc IEEE Signal Processing Magazine}, about 200 refereed scientific papers and book chapters, and three patents. He coedited three books. His 
research 
interests include statistical and compressive signal processing methods and their application to inference and learning problems. 

Dr.\ Hlawatsch was a member of the IEEE SPCOM Technical Committee from 2004 to 2009. He was a Technical Program Co-Chair of EUSIPCO 2004 and served on the technical committees of numerous IEEE conferences. He was an Associate Editor for the {\sc IEEE Transactions on Signal Processing} from 2003 to 2007, the {\sc IEEE Transactions on Information Theory} from 2008 to 2011, and the {\sc IEEE Transactions on Signal and Information Processing over Networks} from 2014 to 2017. He coauthored papers that won an IEEE Signal Processing Society Young Author Best Paper Award and a Best Student Paper Award at IEEE ICASSP 2011. He is a EURASIP Fellow.
\end{IEEEbiography}

\begin{IEEEbiography}[{\includegraphics[width=25mm,height=32.15mm,clip,keepaspectratio]{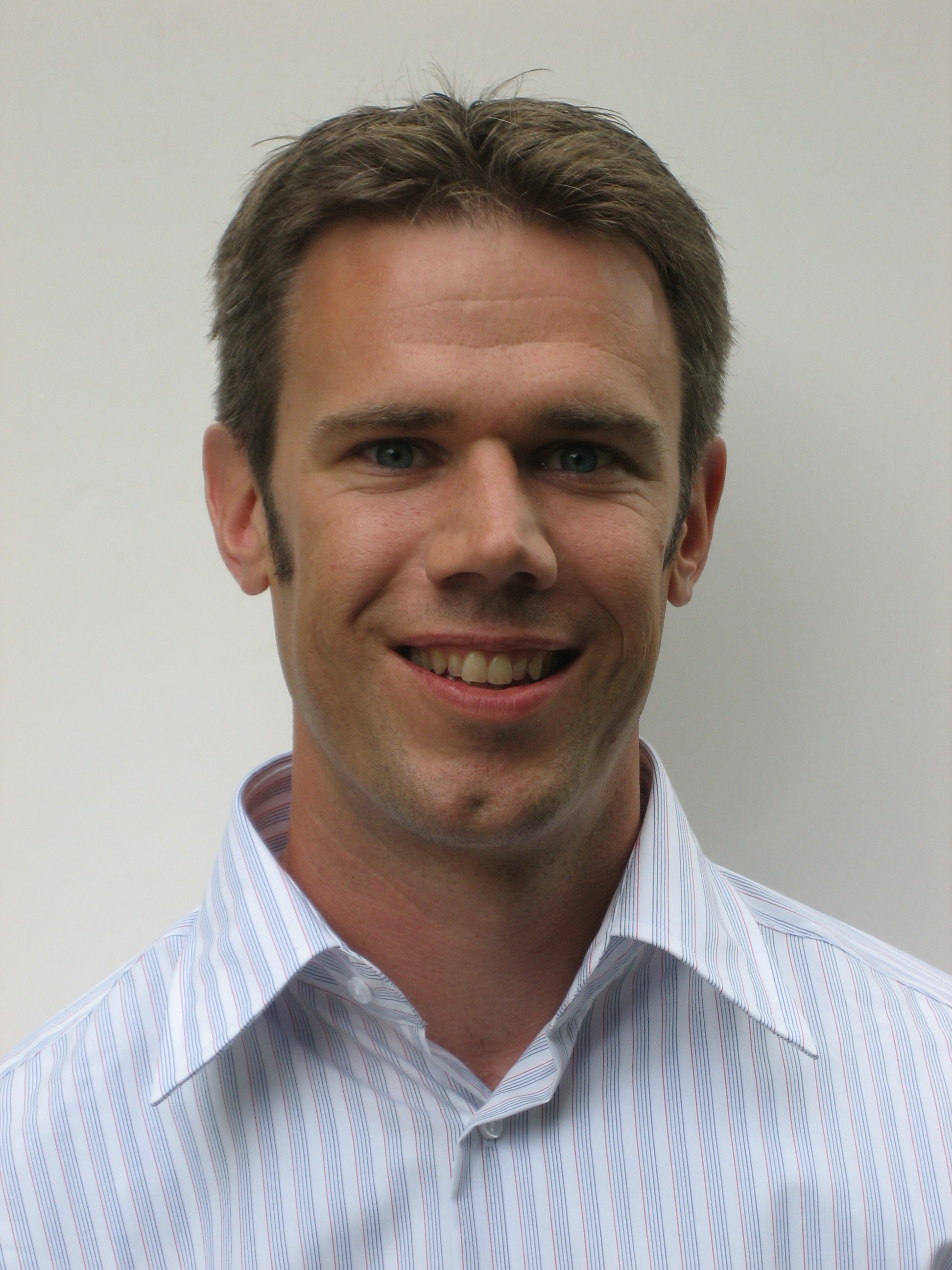}}]{Fredrik~Tufvesson}(S'97--M'04--SM'07--F'17) received his Ph.D.\ in 2000 from Lund University in Sweden. After two years at a startup company, he joined the department of Electrical and Information Technology at Lund University, where he is now professor of radio systems. His main research interest is the interplay between the radio channel and the rest of the communication system with various applications in 5G systems such as massive MIMO, mm wave communication, vehicular communication and radio based positioning.

Fredrik has authored around 90 journal papers and 140 conference papers, he is fellow of the IEEE and recently he got the Neal Shepherd Memorial Award for the best propagation paper in IEEE Transactions on Vehicular Technology and the IEEE Communications Society best tutorial paper award.
\end{IEEEbiography}

\begin{IEEEbiography}[{\includegraphics[width=25mm,height=32.15mm,clip,keepaspectratio]{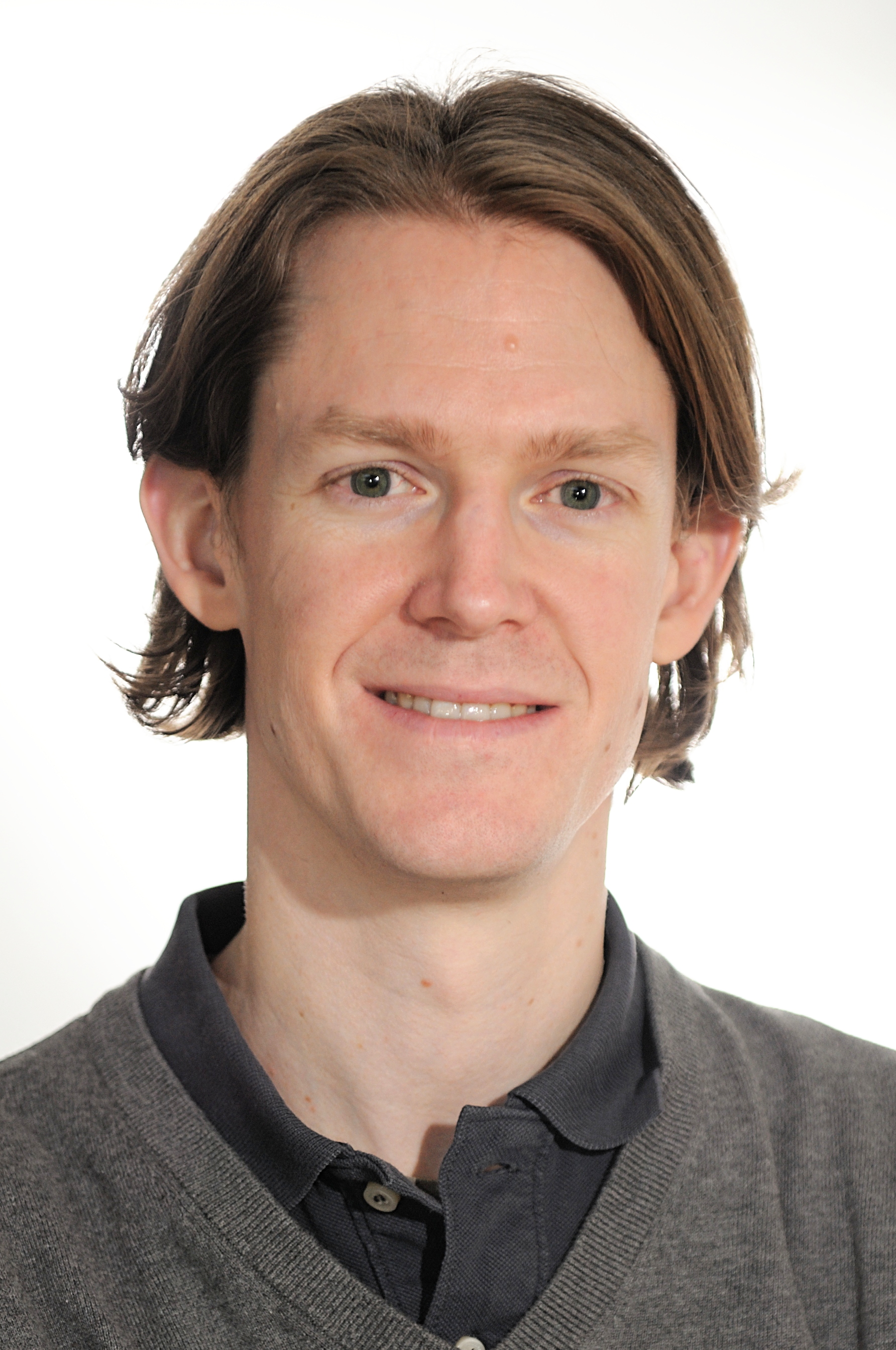}}]{Klaus~Witrisal}(S'98--M'03) received the Ph.D.\ degree from the Delft University of Technology, The Netherlands, in 2002 and the Habilitation degree from the Graz University of Technology, Austria, in 2009, where he is currently an Associate Professor. His research interests are in signal processing for wireless communications, propagation channel modeling, and positioning.

He has been an Associate Editor of the \textsc{IEEE Communications Letters}, a Co-Chair of the TWG ``Indoor'' of the COST Action IC1004 and the EWG ``Localisation and Tracking'' of the COST Action CA15104, a leading Chair of the IEEE Workshop on Advances in Network Localization and Navigation, and TPC (Co)-Chair of the Workshop on Positioning, Navigation and Communication.
\end{IEEEbiography}

\begin{IEEEbiography}[{\includegraphics[width=25mm,height=32.15mm,clip,keepaspectratio]{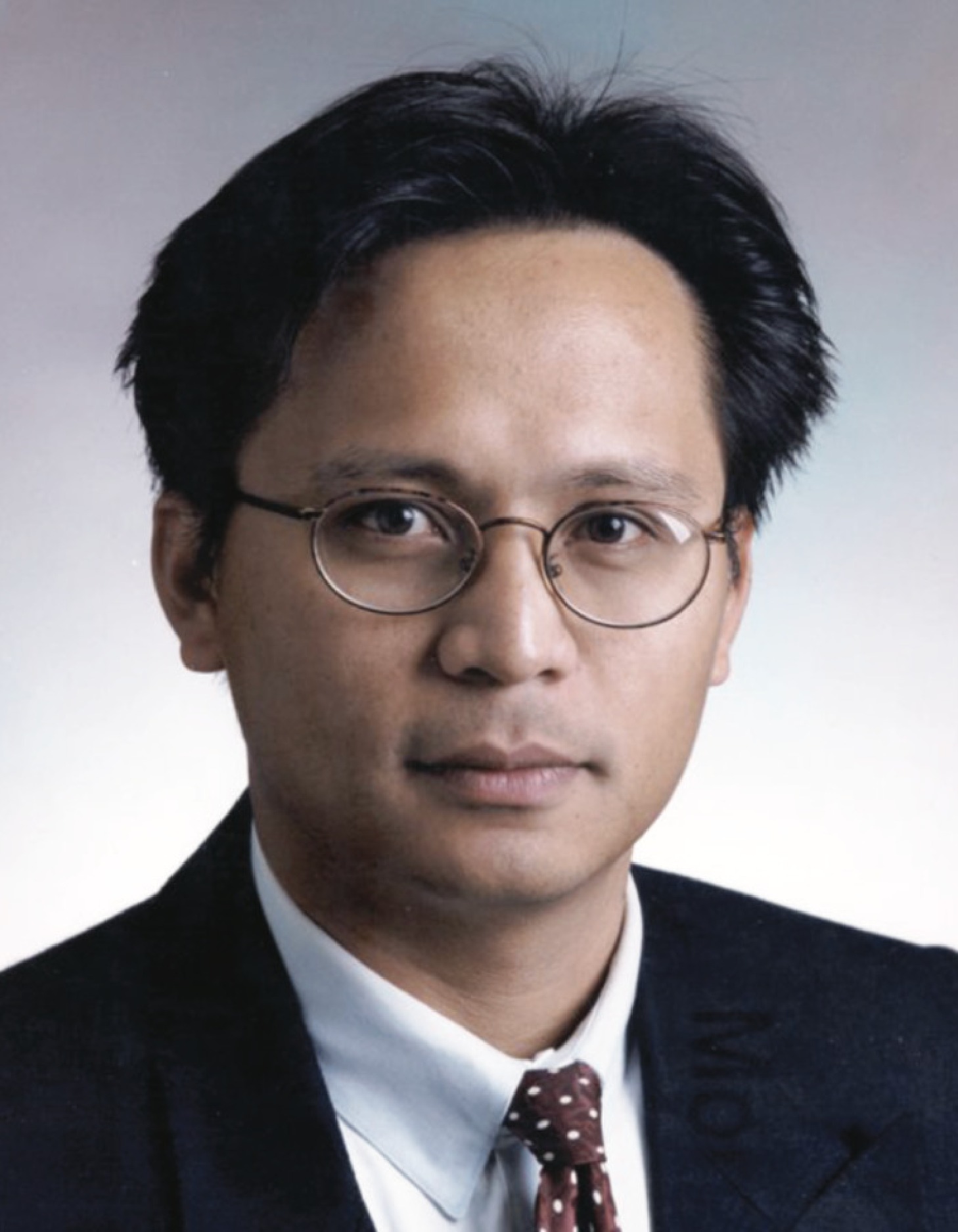}}]{Moe~Z.~Win}(S'85-M'87-SM'97-F'04) is a Professor at the Massachusetts Institute of Technology (MIT) and the founding director of the Wireless Information and Network Sciences Laboratory. Prior to joining MIT, he was with AT\&T Research Laboratories and NASA Jet Propulsion Laboratory.

His research encompasses fundamental theories, algorithm design, and network experimentation for a broad range of real-world problems. His current research topics include network localization and navigation, network interference exploitation, and quantum information science. He has served the IEEE Communications Society as an elected Member-at-Large on the Board of Governors, as elected Chair of the Radio Communications Committee, and as an IEEE Distinguished Lecturer. Over the last two decades, he held various editorial posts for IEEE journals and organized numerous international conferences. Currently, he is serving on the SIAM Diversity Advisory Committee.		

Dr.\ Win is an elected Fellow of the AAAS, the IEEE, and the IET. He was honored with two IEEE Technical Field Awards: the IEEE Kiyo Tomiyasu Award (2011) and the IEEE Eric E. Sumner Award (2006, jointly with R.\ A.\ Scholtz). His publications, co-authored with students and colleagues, have received several awards. Other recognitions include the IEEE Communications Society Edwin H. Armstrong Achievement Award (2016), the International Prize for Communications Cristoforo Colombo (2013), the Copernicus Fellowship (2011) and the {\it Laurea Honoris Causa} (2008) from the Universit\`{a} degli Studi di Ferrara, and the U.S. Presidential Early Career Award for Scientists and Engineers (2004). He is an ISI Highly Cited Researcher.
\end{IEEEbiography}

\end{document}